 \documentclass[final,1p,times]{elsarticle}

\usepackage{amsmath,amssymb,amsfonts}
\usepackage{algorithmic}
\usepackage{graphicx}
\usepackage{textcomp}
\usepackage{todonotes}

\usepackage{tikz}
\usepackage{pgfplots}
\usepgfplotslibrary{colorbrewer}
\pgfplotsset{compat=1.8}
\definecolor{clr1}{RGB}{27,158,119}
\definecolor{clr2}{RGB}{217,95,2}
\definecolor{clr3}{RGB}{117,112,179}
\definecolor{clr4}{RGB}{231,41,138}
\definecolor{clr5}{RGB}{102,166,30}
\definecolor{clr6}{RGB}{230,171,2}
\definecolor{clr7}{RGB}{166,118,29}
\pgfplotsset{
    cycle list={clr1,clr2,clr3,clr4,clr5,clr6,clr7},
}

\usepackage{verbatim}
\usepackage{bbm}    
\newtheorem{proposition}{Proposition}
\newtheorem{corollary}{Corollary}
\newtheorem{remark}{Remark}
\newtheorem{lemma}{Lemma}
\newtheorem{observation}{Observation}
\newtheorem{definition}{Definition}
\newcommand*{\QED}{\hfill\ensuremath{\square}}%

\journal{International Teletraffic Congress ITC 32}

\begin{document}

\begin{frontmatter}

\title{Threshold-based rerouting and replication for\\ resolving job-server affinity relations}

\author[label1]{Youri Raaijmakers\corref{cor1}}
\ead{y.raaijmakers@tue.nl}
\author[label1]{Sem Borst}
\author[label1]{Onno Boxma}

\address[label1]{Department of Mathematics and Computer Science, Eindhoven University of Technology, 5600 MB Eindhoven, The Netherlands}
\cortext[cor1]{Corresponding author}

\begin{abstract}
We consider a system with several job types and two parallel server pools.
Within the pools the servers are homogeneous, but across pools possibly not in the sense
that the service speed of a job may depend on its type as well as the server pool.
Immediately upon arrival, jobs are assigned to a server pool. This could be based on (partial) knowledge of their type, but such knowledge might not be available. 
Information about the job type can however be obtained while the job is in service;
as the service progresses, the likelihood that the service speed of this job type is low increases,
creating an incentive to execute the job on different, possibly faster, server(s). 
Two policies are considered:  reroute the job to the other server pool, or replicate it there.

We determine the effective load per server under both the rerouting and replication policy
for completely unknown as well as partly known job types.   
We also examine the impact of these policies on the stability bound,
and find that the uncertainty in job types may significantly degrade the performance.
For (highly) unbalanced service speeds full replication achieves the largest stability bound
while for (nearly) balanced service speeds no replication maximizes the stability bound.
Finally, we discuss how the use of threshold-based policies can help improve the expected latency for completely or partly unknown job types.
\end{abstract}


\end{frontmatter}

\section{Introduction}
\label{sec: introduction}
This paper considers parallel-processing systems with two heterogeneous server pools,
in which a dispatcher assigns jobs to one of the server pools immediately upon arrival.
However, a job is allowed to be \textit{rerouted} to the other pool when its service
has not yet been completed after a certain amount of processing time.
We also consider an alternative option in which such a job is \textit{replicated} at the other pool.
The jobs are assumed to be of different types;
one job type might, e.g., be fast on a server from pool~$1$ and slow on a server from pool~$2$,
whereas this might be reversed for another job type.
We examine the ability of threshold-based policies to deal with such heterogeneity in service speeds:
if the job types are unknown, or only partly known, can we use a threshold
for the amount of processing time after which the job should be rerouted (or replicated)
in order to improve stability of the system and reduce latency?

Replication schemes as described above were introduced to mitigate the adverse effect of so-called stragglers 
on the system performance, see for example \cite{AKGSLSH-ROMRC,DB-TAS,VGMSRS-LLR}.
Closely related to the present paper is~\cite{APS-ESM} which studies the replication policy
under the assumption of i.i.d.\ replicas and homogeneous servers.
An approximation for the expected latency is derived for both exponential and shifted-exponential job size distributions.
Moreover, the trade-off between the expected latency and the cost,
defined as the sum of the processing times of each replica involved in the job execution, is analyzed via simulation.
Replication schemes are shown to reduce both cost and expected latency in case of heavy-tailed job size distributions.
Also closely related is \cite{J-BSC} which focuses on the throughput-optimal replication policy
under the assumption of identical replicas and a continuous speed distribution at each server.
The Markov Decision Process formulation for the optimal replication policy is in general intractable,
and therefore an upper bound is derived.
In addition, a myopic \textit{MaxRate} policy is introduced, which depends on the number of unfinished jobs
after a certain action and the expected remaining processing time.

In all the above papers the speed variations experienced by the various replicas are essentially assumed
to be purely random in nature.
The multi-type set-up that we consider in the present paper allows for intrinsic differences in speed
across servers depending on the specific characteristics of each individual job.
Such heterogeneity in service speeds captures systematic job-server affinity relations
which may arise from data locality issues but may also reflect soft compatibility constraints
that are increasingly prevalent in data center environments.
We investigate the effectiveness of threshold-based replication and rerouting in the presence of such
underlying job-server affinity relations.

The performance of the threshold-based replication policy is strongly linked
to the stability of redundancy scheduling,
where several replicas are created for each job immediately upon arrival,
see for example \cite{J-ERT,KRW-JRMS,KR-RAGC,RBB-ASR}.
Specifically, it is demonstrated in~\cite{RBB-ASR} that in case of \textit{known} job types
and a probabilistic type-dependent job assignment strategy,
\textit{no replication} maximizes the stability bound
for New-Better-than-Used (NBU) job size distributions. 
In contrast, in case of \textit{unknown} job types,
\textit{full replication} achieves a larger stability bound than no replication
for New-Worse-than-Used (NWU) job size distributions.

Likewise, the threshold-based rerouting policy is closely connected to the problem
of learning the type of a job in an online manner
as considered by \textit{Bimpikis and Markakis}~\cite{BM-LHSS}.
In their model there is a server pool that is compatible with jobs of all types
and a server pool that is either compatible or incompatible with a job depending on its type.
Here (in)compatible means that the service of a job can(not) be fulfilled by a server in this pool.
The job types can be learned via observing the processing time.
Indeed, as time elapses and the service has not been completed,
the likelihood that the job is incompatible with the specific server it is on increases,
and in~\cite{BM-LHSS} the job is therefore rerouted to the other server pool
when the likelihood that the server is incompatible exceeds some value, also called threshold.
For related literature on this learning problem we refer to~\cite{BM-LHSS} and the references therein. 

In this paper we consider similar threshold-based policies as in~\cite{BM-LHSS},
but defined in terms of the received processing time, and extend the model
of~\cite{BM-LHSS} to allow for completely general type-dependent service speed variations.
Furthermore, we include replication as an additional option besides rerouting,
resulting in concurrent execution of replicas at possibly different speeds as in~\cite{RBB-ASR}.
Throughout, we use the term \textit{stability bound} to refer to the maximum arrival rate of jobs for which the effective load per server is smaller than one.
The effective load per server is defined as a measure of the amount of time required to serve all offered jobs. 

The key contributions and insights can be summarized as follows.
\begin{enumerate}
\item We determine the effective load per server under both the rerouting and replication policy.
We show that for unknown job types and (highly) unbalanced service speeds
the largest stability bound is achieved by full replication.
In contrast, for (nearly) balanced service speeds the stability bound is maximized by not replicating at all.
Surprisingly, rerouting or replication does not significantly increase the stability bound
in case of unknown job types, in part because of the variability in the job sizes.
\item We also determine the effective load per server in a scenario with known job sizes.
We find that typically there still is a significant performance loss in terms
of the stability bound compared to the case of known job types.
This implies that the uncertainty in the job types plays a more pertinent role
than the variability in the job sizes.
\item We extend our modeling framework to allow for partly known job types.
We observe that decreasing the uncertainty in the job types increases
the stability bound in a convex manner.
\item We discuss how the use of threshold-based policies can also help improve
the expected latency for completely unknown or partly known job types.
While an exact latency analysis is beyond the scope of this paper,
we provide approximations that show a reduction of the expected latency. 
\end{enumerate}

The remainder of the paper is organized as follows.
In Section~\ref{sec: model description} we provide a detailed model description.
Analytical expressions for the effective load per server under threshold-based rerouting or replication
in the case of completely unknown and partly known job types are derived in Section~\ref{sec: stability unknown job sizes}.
In Section~\ref{sec: stability known job sizes} we characterize the effective load per server in a scenario where the thresholds depend
on the job sizes when these are known in advance.
Section~\ref{sec: numerical results} presents extensive numerical results
which quantify the performance implications due to the uncertainty in job types.
In Section~\ref{sec: improving expected latency} we examine how the use of threshold-based policies
can also help reduce expected latency.
Section~\ref{sec: conclusion} contains conclusions and some suggestions for further research. 

\section{Model description}
\label{sec: model description}

Consider a system with $N$ servers and a dispatcher where jobs arrive at rate $\lambda$.
The servers are divided into two server pools, where $n_{i}$ denotes the number of servers in pool~$i$, with $n_{1}+n_{2}=N$.
The dispatcher assigns jobs immediately upon arrival to one of the two server pools according to a static policy as will be further specified later. Each pool has a central queue that follows a non-idling service discipline, in the sense that the servers always serve jobs when the queue is non-empty. 
We allow the sizes $X_{1},X_{2}$ of a generic job on the server pools to be governed by some joint distribution $F_{X}(x_{1},x_{2})$, where $X_{i}$, $i=1,2$, are each distributed as a generic random variable $X$, but not necessarily independent. 
This covers the extreme scenarios of perfect dependence (identical replicas) and no dependence at all (i.i.d.\ replicas), as previously considered in the literature, as special cases. 
The service speeds $R_{1},R_{2}$ for a given job in the two pools may differ, but within the pools servers have the same speed.
For a particular job on a server in pool $i$, $i=1,2$, with size $x_{i}$, $x_{i}/R_{i}$ represents the execution time.
We allow the service speeds $R_{1},R_{2}$ of a generic job
to be governed by some joint distribution $F_{R}(r_{1},r_{2})$, reflecting possible server heterogeneity and job-server affinity relations, thus covering a broad range of common workload models as special cases. 
Examples are the S\&X model \cite{GHBSW-DSSJS}, a scenario with heterogeneous server speeds and the 'output-queued' flexible server model \cite{S-OROQ}. 

For convenience, we consider the case where the joint distribution $F_{R}(r_{1}, r_{2})$ is discrete, and has mass in a finite number of, say,~$J$ points $(r_{1j},r_{2j})$ with corresponding probabilities $p_{j}$, $j = 1, \dots,J$. This system may equivalently be thought of as having $J$ job types, where $r_{ij}$ is the service speed of type-$j$ jobs at servers in pool~$i$.

\begin{figure}[htbp]
\centering
\includegraphics[width=4cm]{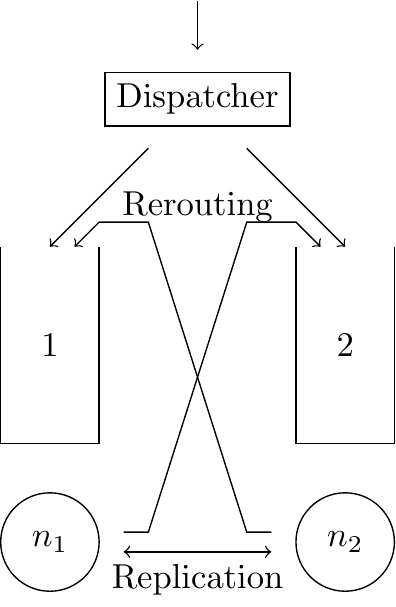}
\caption{Representation of the model that illustrates the difference between the rerouting and the replication policy.}
\label{fig: delay model}
\end{figure}  

As an important feature of our model we consider the following two resource allocation policies: \textit{rerouting} and \textit{replication}, see also Figure~\ref{fig: delay model}.
In both policies, a job exits the system once its service requirement has been fully fulfilled.
However, if the job has received a given amount of processing time (we focus on the case of a fixed service time threshold $\boldsymbol{\tau}=(\tau_{1},\tau_{2})$), then the dispatcher, under the rerouting policy, reroutes the job to the other server pool.
Under the replication policy, the job is replicated in the other server pool while also staying in service at its original pool.
In both policies the service does not carry over, i.e., the entire service requirement should be fulfilled by a server in the `new' pool. 
Thus, the main difference between the policies is that in the case of replication, when the job is in service for an amount of processing time larger than the threshold, there is still a replica in service at the original pool. 

We make the following assumption: replicas of a job after replication have preemptive-resume priority over jobs that are not yet replicated, if they can get simultaneous service at both server pools. If both replicas cannot immediately get simultaneous service, then they wait until one server in each pool is available. This ensures that after replication the replicas will always receive simultaneous service. This assumption additionally helps eschew stability issues that can arise due to local priorities in scenarios with simultaneous resource possession, see for example~\cite[Section~8.4]{KY-SN}.

For equally large server pools, i.e., $n_{1}=n_{2}$, the preemptive-resume priority implies that jobs never have to wait after replication. Indeed, the numbers of servers serving replicated jobs are equal at all times and therefore it can never occur that a job is replicated, while all the servers in the other server pool are serving jobs that are already replicated. 
For unequal sizes of the server pools, i.e., $n_{1} \neq n_{2}$, it might occur that after replication both replicas have to wait before getting service. 

Replicas are abandoned as soon as one finishes service.
The replication policy with threshold $\boldsymbol{\tau}=\boldsymbol{0}$ is referred to as the \textit{full redundancy} policy. For threshold $\boldsymbol{\tau} = \boldsymbol{\infty}$ the rerouting and replication policy are equivalent, and will be called the \textit{zero redundancy} policy.

In this paper we restrict ourselves to the scenario with two server pools. Extension to an arbitrary number of server pools is left as a topic for further research. 
We focus on increasing the achievable stability bound by optimizing threshold values in terms of the received processing times.

\section{Stability for unknown job sizes}
\label{sec: stability unknown job sizes}
In Section~\ref{subsec: stability unknown job sizes} we derive analytical expressions for the effective load per server for both the rerouting and replication policy under a given assignment and threshold values. In Section~\ref{subsec: stability unknown job sizes specific scenarios} we specify these allocation fractions for three cases: Completely unknown job types, partly unknown job types and completely known job types.
Throughout we assume that the job sizes are unknown.

We attach superscripts \textit{Rer} and \textit{Rep} to metrics that correspond to the \textit{rerouting} and \textit{replication} policy, respectively.

\subsection{Load of the system under a given assignment}
\label{subsec: stability unknown job sizes}

Let $A$ denote the $2 \times J$ stochastic assignment matrix with elements $\alpha_{ij}$. Under the $S(A,\boldsymbol{\tau})$ policy we assign a fraction $\alpha_{ij}$ of type-$j$ jobs to server pool~$i$ and reroute (or replicate) at this pool as soon as the processing time equals $\tau_{i}$ for $i=1,2$. 
We define $l:= 3-i$ as the other server pool than $i$, so that $\alpha_{ij}+\alpha_{lj}=1$ for all $j=1,\dots,J$. 

\begin{proposition}
\label{thm: stability TwoW_SF_Rer}
The effective load of server pool~$i$, for $i=1,2$, in the system with rerouting under the $S(A,\boldsymbol{\tau})$ policy is
\begin{align}
\label{eq: stability TwoW_SF_Rer}
\rho_{i,S(A,\boldsymbol{\tau})}^{\text{Rer}} = \frac{\lambda \mathbb{E}\left[B^{\text{Rer}}_{i}(A,\boldsymbol{\tau})\right]}{n_{i}},
\end{align}
where the expected service time requirement of an arbitrary job, assigned either to pool~$i$ or pool~$l$, at pool~$i$ is
\begin{align}
\mathbb{E}\left[B^{\text{Rer}}_{i}(A,\boldsymbol{\tau})\right] &= \sum_{j=1}^{J} p_{j} \alpha_{ij} \mathbb{E}\left[ \min\left\{\frac{X_{i}}{r_{ij}},\tau_{i}\right\}\right] + \sum_{j=1}^{J} p_{j} \alpha_{lj} \mathbb{E}\left[ \frac{X_{i}}{r_{ij}} \mathbbm{1} \left\{ \frac{X_{l}}{r_{lj}} > \tau_{l} \right\} \right]. 
\label{eq: rerouting service requirement pool i}
\end{align}
\end{proposition}
\noindent \textbf{Proof:}
When allocating a job to server pool~$i$, there are two possibilities: i) the job finishes before rerouting; ii) the job is rerouted, in which case the job receives $\tau_{i}$ units of service at this server pool. However, jobs that started in server pool~$l$ can be rerouted to pool~$i$ as well. The proof then follows from noting that $\mathbb{E}\left[ \min\left\{\frac{X_{i}}{r_{ij}},\tau_{i}\right\}\right]$ represents the expected amount of processing of a job in server pool $i$ before completing or being rerouted to server pool $l$, and $\mathbb{E}\left[ \frac{X_{i}}{r_{ij}} \mathbbm{1} \left\{ \frac{X_{l}}{r_{lj}} > \tau_{l} \right\} \right]$ represents the expected received amount of processing of a job in server pool $i$, if any, after having received $\tau_{l}$ units in pool $l$ first.\QED \\

To simplify the expressions for the replication policy, we introduce some notation:
\begin{align*}
&k_{il,j}^{m}(y) = \mathbb{E}\left[\left( \min\left\{\frac{X_{i}}{r_{ij}}-y , \frac{X_{l}}{r_{lj}} \right\} \right)^{m} \mathbbm{1}\left\{\frac{X_{i}}{r_{ij}} > y \right\} \right].
\end{align*}
This represents the $m$-th moment of the amount of time that a job will be processed in both server pools, if any, after being processed up to $y$ time units in server pool $i$ first. In case of $m=1$ the superscript is omitted.

\begin{proposition}
\label{thm: stability TwoW_SF_Red}
The effective load per server in pool~$i$, for $i=1,2$, in the system with replication under the $S(A,\boldsymbol{\tau})$ policy is
\begin{align}
\rho_{i,S(A,\boldsymbol{\tau})}^{\text{Rep}} = \frac{\lambda \mathbb{E}\left[B^{\text{Rep}}_{i}(A,\boldsymbol{\tau})\right]}{n_{i}},
\label{eq: stability TwoW_SF_Red}
\end{align}     
where the expected service time requirement of an arbitrary job, assigned either to pool~$i$ or pool~$l$, at pool~$i$ is
\begin{align}
\mathbb{E}\left[B^{\text{Rep}}_{i}(A,\boldsymbol{\tau})\right] &= \sum_{j=1}^{J} p_{j} \alpha_{ij} \bigg( \mathbb{E}\left[ \min\left\{\frac{X_{i}}{r_{ij}},\tau_{i}\right\}\right] + k_{il,j}(\tau_{i}) \bigg) + \sum_{j=1}^{J} p_{j} \alpha_{lj} k_{li,j}(\tau_{l}). 
\label{eq: replication service requirement pool i}
\end{align}
\end{proposition}
\noindent \textbf{Proof:}
When allocating a job to server pool~$i$, there are two possibilities: i) the job finishes before replication; ii) the job is replicated, in which case the job spends at least $\tau_{i}$ time units in service at this server pool. The (remaining) expected service time requirement of jobs starting at server pool $i$ after replication, if any, is $k_{il,j}(\tau_{i})$. 
However, jobs that started in server pool~$l$ can be replicated to pool~$i$ as well. The (remaining) expected service time requirement of these jobs, if any, is $k_{li,j}(\tau_{l})$. \QED \\

Evidently, $\rho_{i,S(A,\boldsymbol{\tau})}^{\text{Rer}} < 1$ (or $\rho_{i,S(A,\boldsymbol{\tau})}^{\text{Rep}} < 1$), for $i=1,2$, is a necessary condition for stability. It is quite plausible that this condition is in fact also sufficient for the system to be stable (under the preemptive-resume priority policy for replicated jobs). 

Indeed, for multiserver queues it is well known that the system is stable if and only if the load per server is smaller than one, where the arrival process can be quite general (see for example~\cite[Chapter~1]{B-SPQT} or~\cite[Chapter~7]{W-IQN}).  
Moreover, in~\cite[Proposition~7.4.12]{W-IQN} it is proved that, under the assumption that the sequence of inter-arrival and service times of the jobs at the queues is ergodic and stationary, the stability of two $G/G/1$ queues in tandem is independent of the departure process of the first queue. However, in our system both initially assigned \textit{and} rerouted (or replicated) jobs arrive at a server pool. In addition, replicated jobs must receive service simultaneously at both server pools, implying that servers may be idling even when there are jobs waiting. As a result, it is hard to rigorously establish that a load per server smaller than one is sufficient for the system to be stable. 

\begin{remark}
\label{rem: zero redundancy full redundancy policy}
The expected service time requirement at server pool~$i$ for the \textit{zero redundancy} policy is equal to
\begin{align}
\mathbb{E}\left[B^{\text{Rer}}_{i}(A,\boldsymbol{\infty})\right] &= \mathbb{E}\left[B^{\text{Rep}}_{i}(A,\boldsymbol{\infty})\right] = \sum_{j=1}^{J} p_{j} \alpha_{ij} \mathbb{E}\left[ \frac{X_{i}}{r_{ij}}\right],
\label{eq: zero redundancy service requirement pool i}
\end{align}
and the expected service time requirement at server pool~$i$ for the \textit{full redundancy} policy is equal to
\begin{align}
&\mathbb{E}\left[B^{\text{Rep}}_{i}(A,\boldsymbol{0})\right] = \sum_{j=1}^{J} p_{j} \mathbb{E}\left[\min\left\{\frac{X_{i}}{r_{ij}} , \frac{X_{l}}{r_{lj}} \right\} \right].
\label{eq: full redundancy service requirement pool i}
\end{align}
\end{remark}

Observe that the expected service time requirement for the full redundancy policy is independent of the assigned fractions. The achievable stability bound for this policy coincides with that derived in \cite{RBB-ASR}. Furthermore, note that in case of identical replicas the effective load per server for both the zero redundancy and full redundancy policy is insensitive to the job size distribution, given its mean. 

\subsection{Specific assignment fractions}
\label{subsec: stability unknown job sizes specific scenarios}
In this subsection, we specify the assignment fractions in Equations~\eqref{eq: rerouting service requirement pool i} and~\eqref{eq: replication service requirement pool i} for the three cases of completely unknown, partly unknown and completely known job types. Let $q^{i}_{j}$ denote the fraction of type-$j$ jobs that are assigned to server pool $i$ in these three cases. The proofs are straightforward, and hence omitted.

\begin{corollary}
\label{cor: allocations unknown job types}
For the case of unknown job types, Propositions~\ref{thm: stability TwoW_SF_Rer} and~\ref{thm: stability TwoW_SF_Red} hold with
\begin{align}
\alpha_{ij} &= q^{i}, &\text{for } i=1,2 \text{ and } j=1,\dots,J. 
\label{eq: allocation alpha completely unknown}
\end{align}
\end{corollary}

We proceed with the case of partly known job types by which we mean that every arriving job is \textit{believed} to be of a specific type. This can be thought of as a \textit{label} indicating the likely type of a job. 
Let $p_{j \rightarrow j^{*}}$ denote the probability that a job of type $j$ is believed to be of type $j^{*}$, with $\sum_{j^{*}=1}^{J} p_{j \rightarrow j^{*}}=1$. 
The scenario $p_{j \rightarrow j}=1$ corresponds to the case of known job types, whereas $p_{j \rightarrow j^{*}} = p_{j^{*}}$, for all $j=1,\dots,J$, corresponds to the case of unknown job types. 

\begin{corollary}
\label{cor: allocations partly known job types}
For the case of partly known job types, Propositions~\ref{thm: stability TwoW_SF_Rer} and~\ref{thm: stability TwoW_SF_Red} hold with
\begin{align}
\alpha_{ij} &= \sum_{j^{*}=1}^{J} p_{j \rightarrow j^{*}} q_{j^{*}}^{i}, &\text{for } i=1,2 \text{ and } j=1,\dots,J. 
\label{eq: allocation alpha partly known}
\end{align}
\end{corollary}

Observe that in the case of completely unknown job types Equation~\eqref{eq: allocation alpha partly known} becomes 
$\alpha_{ij} = \sum_{j^{*}=1}^{J} p_{j^{*}} q_{j^{*}}^{i}$ for $i=1,2$, which is equivalent to Equation~\eqref{eq: allocation alpha completely unknown} since it is independent of the job types.

\begin{corollary}
\label{cor: allocations completely known job types}
For the case of known job types, Propositions~\ref{thm: stability TwoW_SF_Rer} and~\ref{thm: stability TwoW_SF_Red} hold with
\begin{align}
\alpha_{ij} &= q_{j}^{i}, &\text{for } i=1,2 \text{ and } j=1,\dots,J. 
\label{eq: allocation alpha known}
\end{align}
\end{corollary}

Substituting these assignment fractions in Equation~\eqref{eq: zero redundancy service requirement pool i} gives
\begin{align*}
\mathbb{E}\left[B^{\text{Rer}}_{i}(A,\boldsymbol{\infty})\right] &= \mathbb{E}\left[B^{\text{Rep}}_{i}(A,\boldsymbol{\infty})\right] = \sum_{j=1}^{J} p_{j} q_{j}^{i} \mathbb{E}\left[ \frac{X_{i}}{r_{ij}}\right].
\end{align*}
This coincides with the stability condition derived in~\cite{RBB-ASR} for no replication and known job types.
Moreover, in~\cite{RBB-ASR} it is proved that no replication gives a strictly larger stability bound than replication in the case of NBU job size distributions, independent of the server speeds. For NWU job size distributions examples show that both no replication and full replication can give a larger stability bound depending on the server speeds. Examples in which neither no replication nor full replication gives a larger stability bound have not been found, see~\cite{RBB-ASR}.\\ 

\begin{definition}
\label{def: n-threshold rerouting policy}
A rerouting policy corresponds to a stochastic assignment matrix $A$ with elements $\alpha_{ij}$, where we assign a fraction $\alpha_{ij}$ of type-$j$ jobs to server pool~$i$ and vectors $(\tau_{i1}, \tau_{i2}, \dots, \tau_{iK_i}) \in {\mathbb R}_+^{K_i}$
with possibly $K_i = 0$ or $K_i = \infty$, $i = 1, 2$. Here a job of size~$\boldsymbol{x}$ that is assigned to server pool~$i$, is processed for up to $\tau_{i1}$ time units,
and then successively restarted and processed in server pool~$l$ for up to $\tau_{i2}$ time units,
restarted and processed in server pool~$i$ for up to $\tau_{i3}$ time units, etc. Eventually this job is restarted and processed for up to $\tau_{iK_i}$ time units
in server pool~$i$ or~$l$ if $K_i$ is odd or even, respectively,
and ultimately processed for any amount of time in server pool~$i$ or~$l$ if $K_i$ is even or odd, respectively,
until the job is completed, whichever occurs first (or, as long as the job has not been completed).
\end{definition}

In this section we focused on single threshold policies. The effective load per server for the rerouting policy can also be derived for the case of an arbitrary $n$-threshold policy as defined in Definition~\ref{def: n-threshold rerouting policy}, see~\ref{sec app: learning non threshold policy}. However, this would make the optimization over all the parameters, as done numerically in Section~\ref{sec: numerical results}, much more complex.

\section{Stability for known job sizes}
\label{sec: stability known job sizes}
In the previous section we derived the effective load per server for unknown job sizes. In this section, we consider the case of known job sizes, which is indicated by an additional superscript \textit{KS}. We allow the assignment fractions and the threshold values to depend on the size of the job at the server pool, which is indicated by a superscript $x$. Thus, the threshold at a server pool may differ for every arriving job, whereas in the previous section the threshold at the pool was equal for all jobs. The joint density of the job sizes at the server pools is denoted by $f_{X}(\cdot,\cdot)$.
The effective load per server for the rerouting and replication policy are derived in Propositions~\ref{thm: stability TwoW_SF_Rer_KS} and~\ref{thm: stability TwoW_SF_Red_KS}, respectively.

\begin{proposition}
\label{thm: stability TwoW_SF_Rer_KS}
The effective load per server in pool~$i$, for $i=1,2$, in the system with rerouting under the $S(A^{x},\boldsymbol{\tau}^{x})$ policy, in case of known job sizes, is
\begin{align}
\label{eq: stability TwoW_SF_Rer_KS}
\rho_{i,S(A^{x},\boldsymbol{\tau}^{x})}^{\text{Rer,KS}} = \frac{\lambda \mathbb{E}\left[B^{\text{Rer,KS}}_{i}(A^{x},\boldsymbol{\tau}^{x})\right]}{n_{i}},
\end{align}
where
\begin{align}
&\mathbb{E}\left[B^{\text{Rer,KS}}_{i}(A^{x},\boldsymbol{\tau}^{x})\right] = \iint_{\mathbb{R}^{2}} \left( \sum_{j=1}^{J} p_{j} \alpha_{ij}^{x} \min\left\{\frac{x_{i}}{r_{ij}},\tau^{x}_{i} \right\} + \sum_{j=1}^{J} p_{j} \alpha_{lj}^{x} \frac{x_{i}}{r_{ij}} \mathbbm{1}\left\{\frac{x_{l}}{r_{lj}}>\tau^{x}_{l} \right\} \right) f_{X}(x_{1},x_{2}) \text{d}x_{1} \text{d}x_{2}.
\label{eq: rerouting service requirement pool i KS}
\end{align}
\end{proposition}
\noindent \textbf{Proof:}
The proof follows along the same lines as the proof of Proposition~\ref{thm: stability TwoW_SF_Rer}. \QED

\begin{proposition}
\label{thm: stability TwoW_SF_Red_KS}
The effective load per server in pool~$i$, for $i=1,2$, in the system with replication under the $S(A^{x},\boldsymbol{\tau}^{x})$ policy, in case of known job sizes, is
\begin{align}
\rho_{i,S(A^{x},\boldsymbol{\tau}^{x})}^{\text{Rep,KS}} = \frac{\lambda \mathbb{E}\left[B^{\text{Rep,KS}}_{i}(A^{x},\boldsymbol{\tau}^{x})\right]}{n_{i}},
\label{eq: stability TwoW_SF_Red_KS}
\end{align}     
where
\begin{align}
\mathbb{E}\left[B^{\text{Rep,KS}}_{i}(A^{x},\boldsymbol{\tau}^{x})\right] &= \iint_{\mathbb{R}^{2}} \left( \sum_{j=1}^{J} p_{j} \alpha_{ij}^{x} \bigg( \min\left\{\frac{x_{i}}{r_{ij}},\tau^{x}_{i} \right\} + k^{\text{KS}}_{il,j}(\boldsymbol{x},\tau^{x}_{i})\bigg) \right. \nonumber \\
&\qquad \left. + \sum_{j=1}^{J} p_{j} \alpha_{lj}^{x} k^{\text{KS}}_{li,j}(\boldsymbol{x},\tau^{x}_{l}) \right) f_{X}(x_{1},x_{2}) \text{d}x_{1} \text{d}x_{2}, 
\label{eq: replication service requirement pool i KS}
\end{align}
where $k^{\text{KS}}_{il,j}(\boldsymbol{x},y) =  \min\left\{\frac{x_{i}}{r_{ij}}-y,\frac{x_{l}}{r_{lj}} \right\} \mathbbm{1}\left\{ \frac{x_{i}}{r_{ij}} > y \right\}$.
\end{proposition}
\noindent \textbf{Proof:}
The proof follows along the same lines as the proof of Proposition~\ref{thm: stability TwoW_SF_Red}. \QED

Again, the effective load per server can also be derived for the case of an arbitrary $n$-threshold policy, see~\ref{sec app: learning non threshold policy}.

\begin{definition}
Let $(n_{i1}^{x}, n_{i2}^{x}, \dots, n_{iK_i}^{x}) \in {\mathbb N}^{K_i}$ be vectors,
with $\sum_{k = 1}^{K_i} n_{ik}^{x} \leq J - 1$, $n_{ik}^{x} \geq 1$ for all $k=1, \dots, K_i$ and $i=1,2$.
Given the set $J_{ik-1}$, with $J_{i0} = \{1, 2, \dots, J\}$,
let $r_{ik}$ be the $n_{ik}^{x}$-th highest service speed on server pool~$i(k)$ among the job types in $J_{ik-1}$
and let $J_{ik}$ be the job types in $J_{ik-1}$ that do not have the $n_{ik}^{x}$ highest service speeds on server pool~$i(k)$, where $i(k) = i$ if $k$ is odd and $i(k) = l$ if $k$ is even and $i=1,2$
The set $J_{ik}$ corresponds to the remaining possible job types of a job that is initially assigned to server pool~$i$ and rerouted $k$ times (according to a specific policy). 
Then the rerouting policy that corresponds to some stochastic assignment matrix $A^{x}$
and vectors $(\tau_{i1}^{x}, \tau_{i2}^{x}, \dots, \tau_{iK_i}^{x}) \in {\mathbb R}_+^{K_i}$,
with $\tau_{ik}^{x} = \min_{j \in J_{ik-1}}^{(n_{ik})} x_{i(k)} / r_{i(k)j}$ for $k = 1, \dots, K_i$
and $i=1,2$, where $\min^{(n)}$ denotes the $n$-th order statistic, 
is said to be associated with the vectors $(n_{i1}^{x}, n_{i2}^{x}, \dots, n_{iK_i}^{x}) \in {\mathbb N}^{K_i}$.\\
\end{definition}

Any rerouting policy that is associated with two particular vectors as described in Definition~2 is said to be a rational policy.
Note that a rational policy involves at most $J - 1$ threshold values,
and the next observation provides an intuitive explanation for that.

\begin{observation}
In the case of known job sizes and $J$ job types, for each job there are at most $J$ time points, referred to as rational time points, at which the job could possibly fulfill its service requirement, which depend on the job size, server speeds and the threshold values. Moreover, we (only) gain information about the job type at these (rational) time points.
\end{observation}

The next lemma establishes that in order to achieve maximum stability for known job sizes, we may restrict within the class of all rerouting policies as specified in Definition~1, to the subclass of rational policies as defined in Definition~2.

\begin{lemma}
For any arbitrary rerouting policy there exists a rational rerouting policy that is better in the sense that for a job of any size,
given the initial assignment to one of the server pools,
it uses at most the same cumulative amount of processing time in each of the server pools.
\label{lem: threshold values rerouting}
\end{lemma}
\textbf{Proof by induction in the number of remaining possible job types:}\\
\noindent \underline{Base case:} Suppose that for a job initially assigned to server pool~$i$, the rerouting policy is rational up to (measured in processing time) $t_{J-1}=\min_{j \in J_{iK_i-1}}^{(1)} x_{i(K_i-1)} / r_{i(K_i-1)j}$ time units, at which we know that the job is of the one remaining job type and without rerouting the job would finish at $t_{J}$. Then rerouting \textit{exactly at} $t_{J-1}$ uses at most the same cumulative amount of processing time in each of the server pools as rerouting \textit{between} $t_{J-1}$ and $t_{J}$.\\
\noindent \underline{Inductive step:} Show that for any $m \geq 1$, if the lemma holds for $m$ remaining possible job types, then it also holds for $m+1$ remaining possible job types. 

Suppose that for a job initially assigned to server pool~$i$, the rerouting policy is rational up to (measured in processing time) $t_{J-m-1}=\min_{j \in J_{ik}}^{(n_{ik})} x_{i(k)} / r_{i(k)j}$ time units, with $k$ (the number of reroutings after which the job can possibly be of the $m+1$ remaining job types) such that $|J_{ik}|=m+1-n_{ik}$, and without rerouting the first time the job could possibly finish is $t_{J-m}$ (if it is of a specific type). Note that at $t_{J-m-1}$ we know that the job is of one of the $m+1$ remaining job types. Then rerouting \textit{exactly at} $t_{J-m-1}$ uses at most the same cumulative amount of processing time in each of the server pools as rerouting \textit{between} $t_{J-m-1}$ and $t_{J-m}$. In case of no rerouting, if the job does not finish at $t_{J-m}$ we know that the job is of one of the $m$ remaining job types, and we can apply the induction hypothesis. In case of rerouting at $t_{J-m-1}$ the first time the job could possibly finish, say $t^{*}_{J-m}$, may differ from $t_{J-m}$, however we will never reroute again before $t^{*}_{J-m}$. At this time we can apply the induction hypothesis. \QED

Observe that Lemma~1 proves that there exists a rational rerouting policy that is better than any arbitrary policy given the initial assignment to one of the server pools. To achieve the maximum achievable stability bound we also need to find the appropriate initial assignment matrix $A^{x}$. 

\begin{remark}
For the single-threshold rerouting policy, analyzed in Proposition~3, Lemma~\ref{lem: threshold values rerouting} holds as well. For the optimization we hence have $J-1$ candidate values for the threshold. 
\end{remark}

\section{Numerical results}
\label{sec: numerical results}

In Sections~\ref{sec: stability unknown job sizes} and~\ref{sec: stability known job sizes} we derived the effective load per server for various cases. We now present numerical results to get further insight in the performance implications due to the uncertainty in job types. Throughout this section we denote the policies as follows: Rerouting (Rer), Replication (Rep), Zero redundancy (ZRed) and Full redundancy (FRed). The expected service time requirement for these four policies is respectively given by~\eqref{eq: rerouting service requirement pool i}, \eqref{eq: replication service requirement pool i}, \eqref{eq: zero redundancy service requirement pool i} and \eqref{eq: full redundancy service requirement pool i}.
For known job sizes (KS) we only show the maximum of the rerouting and replication policy, where the expected service requirement is respectively given by \eqref{eq: rerouting service requirement pool i KS} and \eqref{eq: replication service requirement pool i KS}. Surprisingly, in all scenarios that we considered the maximum was achieved by the rerouting policy with known job sizes.

In Section~\ref{subsec: numerical results completely unknown job types} we examine the scenario of completely unknown job types and in Section~\ref{subsec: numerical results partly known job types} that of partly known job types. In both subsections we distinguish between identical and i.i.d.\ replicas with exponentially distributed job sizes with mean $N$. We refer to Appendix~\ref{app sec: additional numerical results} for results where the job sizes are Pareto Type I distributed with minimum possible value $1$ and index $\frac{N}{N-1}$, thus again mean $N$, to ensure that the stability condition for known job types is $\lambda<1$.

The scenario $I=2$, $J=2$, $N=10$, $n_{1}=n_{2}=5$, $(r_{11},r_{21})=(1,r_{\text{slow}})$ and $(r_{12},r_{22})=(r_{\text{slow}},1)$ is considered throughout. All figures show the maximum achievable stability bound denoted by $\lambda_{\text{max}}$ as function of some relevant system parameter. For the maximization of the achievable stability bound, we used a brute-force search with a certain fine multidimensional grid of starting points. 

\subsection{Completely unknown job types}
\label{subsec: numerical results completely unknown job types}

\subsubsection{Identical replicas}
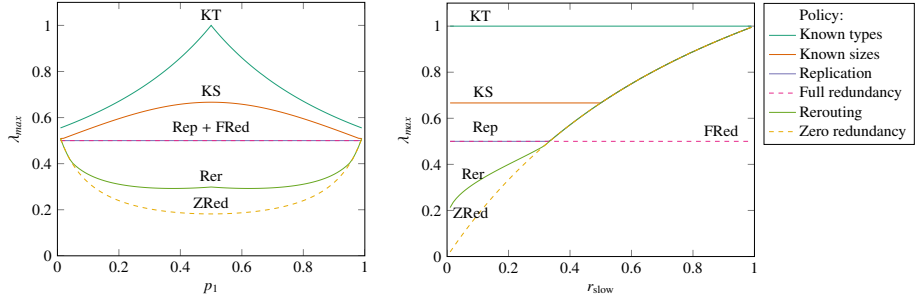
\begin{figure}[htbp]
  \centering
\resizebox{0.9\textwidth}{!}{\newcommand{\dataFigureA}{TikZFigures/StabilityDR_R_TW_SF_Exp_N10_p05_rslow01.csv}
\newcommand{\dataFigureB}{TikZFigures/StabilityDR_R_TW_SF_Exp_N10_p05.csv}

\begin{tabular}{@{}cc@{}}
\begin{tikzpicture}
		\begin{axis}[
			xlabel=$p_{1}$,
			ylabel=$\lambda_{max}$,
			ymin=0,
			ymax=1.1,
			xmin=0,
			xmax=1,
			no markers]
		\addlegendimage{empty legend}
		\addplot+ table [x=prob, y=KT, col sep=comma]{\dataFigureA}
		node[black,pos=0.5,above]{KT};	
		\addplot+ table [x=prob, y=KS, col sep=comma]{\dataFigureA} 							node[black,pos=0.5,above]{KS};
		\addplot+ table [x=prob, y=Rep, col sep=comma]{\dataFigureA};
		\addplot+[dashed] table [x=prob, y=Red, col sep=comma]{\dataFigureA}
		node[black,pos=0.5,above]{Rep + FRed};
		\addplot+ table [x=prob, y=Rer, col sep=comma]{\dataFigureA}
		node[black,pos=0.5,above]{Rer};
		\addplot+[dashed] table [x=prob, y=DN, col sep=comma]{\dataFigureA}
		node[black,pos=0.5,above]{ZRed};		
		\end{axis}
\end{tikzpicture}
&
\begin{tikzpicture}
		\begin{axis}[
			xlabel=$r_{\text{slow}}$,
			ylabel=$\lambda_{max}$,
			ymin=0,
			ymax=1.1,
			xmin=0,
			xmax=1,
			no markers,
			legend pos= outer north east,
			legend cell align=left]
		\addlegendimage{empty legend}
		\addplot+ table [x=rslow, y=KT, col sep=comma]{\dataFigureB}
		node[black,pos=0.1,above,align=left]{KT};	
		\addplot+ table [x=rslow, y=KS, col sep=comma]{\dataFigureB}
		node[black,pos=0.1,above,align=left]{KS};	
		\addplot+ table [x=rslow, y=Rep, col sep=comma]{\dataFigureB}
		node[black,pos=0.1,above,align=left]{Rep};	
		\addplot+[dashed] table [x=rslow, y=Red, col sep=comma]{\dataFigureB}
		node[black,pos=0.9,above,align=left]{FRed};	
		\addplot+ table [x=rslow, y=Rer, col sep=comma]{\dataFigureB}
		node[black,pos=0.1,above,align=left]{Rer};	
		\addplot+[dashed] table [x=rslow, y=DN, col sep=comma]{\dataFigureB}
		node[black,pos=0.1,above,align=left]{ZRed};			

		\addlegendentry{Policy:}
		\addlegendentry{Known types}
		\addlegendentry{Known sizes}
		\addlegendentry{Replication}
		\addlegendentry{Full redundancy}
		\addlegendentry{Rerouting}
		\addlegendentry{Zero redundancy}
		\end{axis}
\end{tikzpicture}
\end{tabular}}
\caption{Achievable stability bound for identical replicas with exponentially distributed job sizes in the scenario $I=2$, $J=2$, $N=10$, $n_{1}=n_{2}=5$, $(r_{11},r_{21})=(1,r_{\text{slow}})$ and $(r_{12},r_{22})=(r_{\text{slow}},1)$ with corresponding probabilities $p_{1}$ and $p_{2}=1-p_{1}$, with $r_{\text{slow}}=0.1$ (left) and $p_{1}=p_{2}=0.5$ (right). In the left sub-figure, the replication and full redundancy policy are overlapping.}
\label{fig: identical delay stability TwoW_SF Exp N10 p05}
\end{figure}

In Figure~\ref{fig: identical delay stability TwoW_SF Exp N10 p05} the maximum achievable stability bound for the various policies is depicted when varying the parameters $p_{1}$ and $r_{\text{slow}}$. Observe that the latter scenario is completely symmetric and therefore $q^{1}=q^{2}=0.5$ is the optimal allocation.

The left sub-figure in Figure~\ref{fig: identical delay stability TwoW_SF Exp N10 p05} shows that the replication policy outperforms the rerouting policy. This generally holds for scenarios where $r_{\text{slow}}$ is relatively small (unbalanced server speeds). The reason is that the rerouting policy becomes unstable, i.e., $\lambda_{\text{max}} \downarrow 0$, as $r_{\text{slow}} \downarrow 0$. In particular, in this scenario, the probability of rerouting the job to an (almost) incompatible server pool, after which we cannot reroute the job again, is strictly positive.

The right sub-figure in Figure~\ref{fig: identical delay stability TwoW_SF Exp N10 p05} reveals that both the rerouting and replication policy achieve the same achievable stability bound for $r_{\text{slow}} > 0.4$, which is the same as for the zero redundancy. Therefore, the achievable stability bound can be achieved by the threshold $\boldsymbol{\tau} = \boldsymbol{\infty}$, see \eqref{eq: zero redundancy service requirement pool i}. So here the threshold does not increase the achievable stability bound. This generally holds in scenarios where $r_{\text{slow}}$ is relatively large (balanced server speeds).
For unbalanced server speeds, the replication policy is equivalent to the full redundancy policy and both outperform the rerouting policy.\\

The achievable stability bound in Figure~\ref{fig: identical delay stability TwoW_SF Exp N10 p05} fails to give information about the processing time per job that is needed to distinguish the job types. In particular, consider the example where one job type has sizes $\epsilon << 1$ and $K >>1$ on a server from pool $1$ or $2$, respectively, and another job type has sizes $K$ and $\epsilon$ on these server pools. In the rerouting policy, after only $\epsilon$ time units we know the job type and we reroute when the service requirement has not been fully fulfilled. Thus, we can distinguish the job types really fast. However, the service time requirement of a rerouted job is in total $2 \epsilon$, while in the case of known types the service requirement of all jobs is $\epsilon$. Therefore, for unbalanced server speeds, the rerouting policy always has a performance loss of at least $33\%$, despite the fact that it can distinguish the job type fast. 

\subsubsection{Independent and identically distributed replicas}
\label{subsec: numerical results independent replicas completely unknown job types}
 
\begin{figure}[htbp]
  \centering
  \resizebox{0.5\textwidth}{!}{\newcommand{\dataFigure}{TikZFigures/StabilityDR_R_TW_SF_Exp_IID_N10_p05.csv}

\begin{tikzpicture}
		\begin{axis}[
			xlabel=$r_{\text{slow}}$,
			ylabel=$\lambda_{max}$,
			ymin=0,
			ymax=2.1,
			xmin=0,
			xmax=1,
			no markers,
			legend pos= outer north east,
			legend cell align=left]
		\addlegendimage{empty legend}
		\addplot+ table [x=rslow, y=KT, col sep=comma]{\dataFigure}
		node[black,pos=0.1,above]{KT};
		\addplot+ table [x=rslow, y=KS, col sep=comma]{\dataFigure}
		node[black,pos=0.05,above]{KS};
		\addplot+ table [x=rslow, y=Rep, col sep=comma]{\dataFigure};
		\addplot+[dashed] table [x=rslow, y=Red, col sep=comma]{\dataFigure}
		node[black,pos=0.15,above]{Rep + FRed};
		\addplot+ table [x=rslow, y=Rer, col sep=comma]{\dataFigure}
		node[black,pos=0.15,below]{Rer};	
		\addplot+[dashed] table [x=rslow, y=DN, col sep=comma]{\dataFigure}
		node[black,pos=0.15,below]{ZRed};		

		\addlegendentry{Policy:}
		\addlegendentry{Known types}
		\addlegendentry{Known sizes}
		\addlegendentry{Replication}
		\addlegendentry{Full redundancy}
		\addlegendentry{Rerouting}
		\addlegendentry{Zero redundancy}
		\end{axis}
\end{tikzpicture}}
\caption{Achievable stability bound for i.i.d.\ replicas with exponentially distributed job sizes in the scenario $I=2$, $J=2$, $N=10$, $n_{1}=n_{2}=5$, $(r_{11},r_{21})=(1,r_{\text{slow}})$ and $(r_{12},r_{22})=(r_{\text{slow}},1)$ with corresponding probabilities $p_{1}=p_{2}=0.5$. The replication and full redundancy policy are overlapping.}
\label{fig: iid delay stability TwoW_SF Exp N10 p05}
\end{figure}
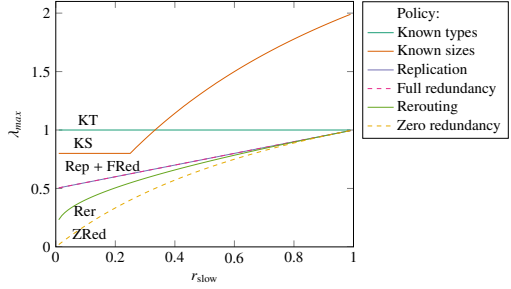

In Figure~\ref{fig: iid delay stability TwoW_SF Exp N10 p05} we consider the same scenario as in Figure~\ref{fig: identical delay stability TwoW_SF Exp N10 p05}, but now for i.i.d.\ replicas.     

Figure~\ref{fig: iid delay stability TwoW_SF Exp N10 p05} shows that the replication policy outperforms the rerouting policy, especially in scenarios with unbalanced server speeds. In this figure, the replication policy is equivalent with the full redundancy policy. This means that learning the job types for the replication policy does not improve the achievable stability bound. Moreover, it can be seen that observing the job sizes can significantly increase the achievable stability bound. For balanced server speeds, rerouting in the case of known sizes even outperforms the policy with known job types. 

\subsection{Partly known job types}
\label{subsec: numerical results partly known job types}
In this section we present numerical results for the achievable stability bound in the case where job types are partly known, i.e., for an arriving job there is a belief that it is of a specific type.

\subsubsection{Identical replicas}
\begin{figure*}[htbp]
\centering
\resizebox{0.9\textwidth}{!}{\newcommand{\dataFigureA}{TikZFigures/StabilityDR_R_TW_SF_Exp_N10_p05_rslow01_likely.csv}
\newcommand{\dataFigureB}{TikZFigures/StabilityDR_R_TW_SF_Exp_N10_p05_rslow06_likely.csv}

\begin{tabular}{@{}cc@{}}
\begin{tikzpicture}
		\begin{axis}[
			xlabel=$p_{1 \rightarrow 1}$,
			ylabel=$\lambda_{max}$,
			ymin=0,
			ymax=1.1,
			xmin=0,
			xmax=1,
			no markers]
		\addplot+ table [x=likely, y=KT, col sep=comma]{\dataFigureA}
		node[black,pos=0.5,above]{KT};
		\addplot+ table [x=likely, y=KS, col sep=comma]{\dataFigureA}
		node[black,pos=0.5,above]{KS};
		\addplot+ table [x=likely, y=Rep, col sep=comma]{\dataFigureA};
		\addplot+[dashed] table [x=likely, y=Red, col sep=comma]{\dataFigureA}
		node[black,pos=0.5,above]{Rep + FRed};
		\addplot+ table [x=likely, y=Rer, col sep=comma]{\dataFigureA}
		node[black,pos=0.5,above]{Rer};
		\addplot+[dashed] table [x=likely, y=DN, col sep=comma]{\dataFigureA}
		node[black,pos=0.5,above]{ZRed};		

		\end{axis}
\end{tikzpicture}
&
\begin{tikzpicture}
		\begin{axis}[
			xlabel=$p_{1 \rightarrow 1}$,
			ylabel=$\lambda_{max}$,
			ymin=0,
			ymax=1.1,
			xmin=0,
			xmax=1,
			no markers,
			legend pos= outer north east,
			legend cell align=left]
		\addlegendimage{empty legend}
		\addplot+ table [x=likely, y=KT, col sep=comma]{\dataFigureB}
		node[black,pos=0.5,above]{KT};
		\addplot+ table [x=likely, y=KS, col sep=comma]{\dataFigureB};
		\addplot+ table [x=likely, y=Rep, col sep=comma]{\dataFigureB};
		\pgfplotsset{cycle list shift=1}
		\addplot+ table [x=likely, y=Rer, col sep=comma]{\dataFigureB};
		\addplot+[dashed] table [x=likely, y=DN, col sep=comma]{\dataFigureB}
		node[black,pos=0.5,below]{KS + Rep + Rer + ZRed};
		\addplot+[clr4,dashed] table [x=likely, y=Red, col sep=comma]{\dataFigureB} node[black,pos=0.5,above]{FRed};		

		\addlegendentry{Policy:}
		\addlegendentry{Known types}
		\addlegendentry{Known sizes}
		\addlegendentry{Replication}
		\addlegendentry{Rerouting}
		\addlegendentry{Zero redundancy}
		\addlegendentry{Full redundancy}
		\end{axis}
\end{tikzpicture}
\end{tabular}}
\caption{Achievable stability bound for identical replicas with exponentially distributed job sizes in the scenario $I=2$, $J=2$, $N=10$, $n_{1}=n_{2}=5$, $(r_{11},r_{21})=(1,r_{\text{slow}})$ and $(r_{12},r_{22})=(r_{\text{slow}},1)$ with corresponding probabilities $p_{1}=p_{2}=0.5$, with $r_{\text{slow}}=0.1$ (left) and $r_{\text{slow}}=0.6$ (right) when varying the probability $p_{1 \rightarrow 1}=p_{2 \rightarrow 2}$.}
\label{fig: identical delay stability TwoW_SF Exp N10 rslow01 rslow06 likely}
\end{figure*}
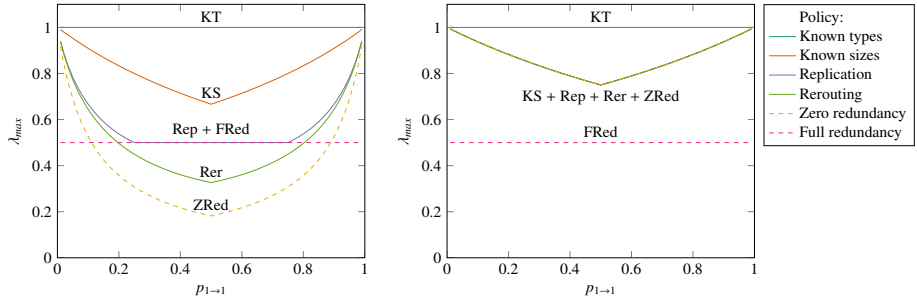

We consider a scenario with unbalanced server speeds and balanced server speeds (Figure~\ref{fig: identical delay stability TwoW_SF Exp N10 rslow01 rslow06 likely}), when varying the probability $p_{1 \rightarrow 1}=p_{2 \rightarrow 2}$. 

Figure~\ref{fig: identical delay stability TwoW_SF Exp N10 rslow01 rslow06 likely} indicates that decreasing the uncertainty in the job types increases the achievable stability bound in a convex manner. Moreover, for unbalanced server speeds, decreasing the uncertainty about the job types at first does not have any effect on the achievable stability bound for the replication policy. However, the replication policy still achieves a larger achievable stability bound than the rerouting policy, especially in scenarios with high uncertainty about the job types.

For balanced server speeds, it can be seen that the achievable stability bounds for the known sizes, replication, rerouting and zero redundancy policy are all equal. Hence, for balanced server speeds, thresholds do not increase the achievable stability bound even if the uncertainty in the job types is decreased. 

\subsubsection{Independent and identically distributed replicas}

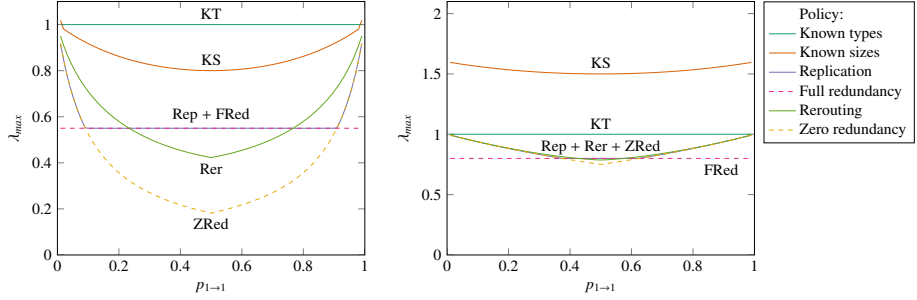
\begin{figure*}[htbp]
\centering
\resizebox{0.9\textwidth}{!}{\newcommand{\dataFigureA}{TikZFigures/StabilityDR_R_TW_SF_Exp_IID_N10_p05_rslow01_likely.csv}
\newcommand{\dataFigureB}{TikZFigures/StabilityDR_R_TW_SF_Exp_IID_N10_p05_rslow06_likely.csv}

\begin{tabular}{@{}cc@{}}
\begin{tikzpicture}
		\begin{axis}[
			xlabel=$p_{1 \rightarrow 1}$,
			ylabel=$\lambda_{max}$,
			ymin=0,
			ymax=1.1,
			xmin=0,
			xmax=1,
			no markers]
		\addplot+ table [x=likely, y=KT, col sep=comma]{\dataFigureA}
		node[black,pos=0.5,above]{KT};
		\addplot+ table [x=likely, y=KJ, col sep=comma]{\dataFigureA}
		node[black,pos=0.5,above]{KS};
		\addplot+ table [x=likely, y=Rep, col sep=comma]{\dataFigureA};
		\addplot+[dashed] table [x=likely, y=Red, col sep=comma]{\dataFigureA}
		node[black,pos=0.5,above]{Rep + FRed};
		\addplot+ table [x=likely, y=Rer, col sep=comma]{\dataFigureA}
		node[black,pos=0.5,below]{Rer};		
		\addplot+[dashed] table [x=likely, y=DN, col sep=comma]{\dataFigureA}
		node[black,pos=0.5,below]{ZRed};		
		\end{axis}
\end{tikzpicture}
&
\begin{tikzpicture}
		\begin{axis}[
			xlabel=$p_{1 \rightarrow 1}$,
			ylabel=$\lambda_{max}$,
			ymin=0,
			ymax=2.1,
			xmin=0,
			xmax=1,
			no markers,
			legend pos= outer north east,
			legend cell align=left]
		\addlegendimage{empty legend}
		\addplot+ table [x=likely, y=KT, col sep=comma]{\dataFigureB}
		node[black,pos=0.5,above]{KT};
		\addplot+ table [x=likely, y=KS, col sep=comma]{\dataFigureB}
		node[black,pos=0.5,above]{KS};
		\addplot+ table [x=likely, y=Rep, col sep=comma]{\dataFigureB};
		\addplot+[dashed] table [x=likely, y=Red, col sep=comma]{\dataFigureB}
		node[black,pos=0.9,below]{FRed};
		\addplot+ table [x=likely, y=Rer, col sep=comma]{\dataFigureB}
		node[black,pos=0.5,above]{Rep + Rer + ZRed};
		\addplot+[dashed] table [x=likely, y=DN, col sep=comma]{\dataFigureB};
		
		\addlegendentry{Policy:}
		\addlegendentry{Known types}
		\addlegendentry{Known sizes}
		\addlegendentry{Replication}
		\addlegendentry{Full redundancy}
		\addlegendentry{Rerouting}
		\addlegendentry{Zero redundancy}
		\end{axis}
\end{tikzpicture}
\end{tabular}}
\caption{Achievable stability bound for i.i.d.\ replicas with exponentially distributed job sizes in the scenario $I=2$, $J=2$, $N=10$, $n_{1}=n_{2}=5$, $(r_{11},r_{21})=(1,r_{\text{slow}})$ and $(r_{12},r_{22})=(r_{\text{slow}},1)$ with corresponding probabilities $p_{1}=p_{2}=0.5$, with $r_{\text{slow}}=0.1$ (left) and $r_{\text{slow}}=0.6$ (right) when varying the probability $p_{1 \rightarrow 1}=p_{2 \rightarrow 2}$.}
\label{fig: independent delay stability TwoW_SF N10 rslow01 rslow06 likely}
\end{figure*}

Figure~\ref{fig: independent delay stability TwoW_SF N10 rslow01 rslow06 likely} reveals that the use of thresholds does not improve the achievable stability bound for the replication policy. Interestingly, for the replication policy in a scenario with (highly) unbalanced server speeds, decreasing the uncertainty has no effect on the achievable stability bound at first. Namely, in the figure it can be seen that the achievable stability bound is constant for $p_{1 \rightarrow 1}$ between $0.1$ and $0.9$, i.e., where the replication and full redundancy policy coincide.

\section{Reducing the expected latency}
\label{sec: improving expected latency}

In the previous sections we focused on the achievable stability bound as key performance metric of interest.
The present section aims to take a first step towards investigating how the use of thresholds can help reduce the expected latency. Throughout this section we assume that the service discipline of the two server pools is FCFS, and that there is only one server per pool, i.e., $n_{1}=n_{2}=1$.

The expected latency has been studied from a somewhat related learning perspective before. Among the numerous studies in the literature, we will highlight two. In \cite{GS-SDHS} the performance of several policies is compared in the setting with heterogeneous server speeds. In particular, the JSQ-DAS (JSQ-$d$ with Accomplishment Sampling) policy uses the idea of server accomplishment to decide which servers to poll, which can be viewed as a learning scheme. In~\cite{AGSS-ESM} a system is introduced that improves the expected latency. The improvement is achieved by replication of small jobs, which ensures that these jobs do not have to wait for large jobs, called stragglers. Learning the optimal threshold for replication is of vital importance for the performance, i.e., the expected latency. \\

Consider the same model as for the analysis of the effective load per server. Again jobs are assigned to a certain server pool and can be rerouted (or replicated) to the other pool after a certain processing time.
In this section we assume that the jobs arrive at the system according to a Poisson process with parameter $\lambda$. We even know what fraction of jobs is rerouted (or replicated) at server pool $i$, namely $p^{\tau}_{i} := \sum_{j=1}^{J} p_{j} \alpha_{ij} \mathbb{P}\left( \frac{X_{i}}{r_{ij}}> \tau_{i}\right)$. By the Poisson thinning property these arrivals follow a Poisson process with parameter $\lambda p^{\tau}_{i}$. However, the departure process, and thus the arrival process at the other server pool after rerouting (or replication), is a complicated process. It is not Poisson, not even a renewal process, even if the job sizes are exponentially distributed. Still, it turns out that we can approximate the mean latency of jobs quite accurately by assuming that jobs are rerouted (or replicated) according to a Poisson process. The Poisson assumption is exact in the extreme cases of $\boldsymbol{\tau}=\boldsymbol{0}$, in which case all jobs are immediately rerouted (or replicated) and of $\boldsymbol{\tau}=\boldsymbol{\infty}$, in which case there is no rerouting (or replication). For large $\boldsymbol{\tau}$, rerouted traffic should also be reasonably close to Poisson traffic, while its contribution to the mean latency is quite small. The Poisson assumption allows us to use the well-known Pollaczek-Khinchin formula for the mean delay in an $M/G/1$ queue. 

The expected latency of a job that is initially assigned to server pool $i$ under the $S(A,\boldsymbol{\tau})$ rerouting policy is given by
\begin{align*}
&\mathbb{E}\left[T_{i}^{\text{Rer}}(A,\boldsymbol{\tau})\right] \approx \frac{\lambda \mathbb{E}\left[\left(B^{\text{Rer}}_{i}(A,\boldsymbol{\tau})\right)^{2}\right]}{2\left(1- \lambda \mathbb{E}\left[B^{\text{Rer}}_{i}(A,\boldsymbol{\tau})\right]\right)} + \mathbb{E}\left[X^{\text{Rer}}_{i}(\boldsymbol{\tau})\right] + \frac{\lambda \mathbb{E}\left[\left(B^{\text{Rer}}_{l}(A,\boldsymbol{\tau})\right)^{2}\right]}{2\left(1- \lambda \mathbb{E}\left[B^{\text{Rer}}_{l}(A,\boldsymbol{\tau})\right]\right)} \sum_{j=1}^{J} p_{j} \mathbb{P}\left( \frac{X_{i}}{r_{ij}}> \tau_{i}\right),
\end{align*}
where, for $i=1,2$, the expected service time requirement is given by Equation~\eqref{eq: rerouting service requirement pool i} and
\begin{align*}
\mathbb{E}\left[ \left( B^{\text{Rer}}_{i}(A,\boldsymbol{\tau})\right)^{2} \right] &= \sum_{j=1}^{J} p_{j} \alpha_{ij} \mathbb{E}\left[ \left( \min\left\{\frac{X_{i}}{r_{ij}},\tau_{i}\right\} \right)^{2} \right] + \sum_{j=1}^{J} p_{j} \alpha_{lj} \mathbb{E}\left[ \left( \frac{X_{i}}{r_{ij}} \right)^{2} \mathbbm{1} \left\{ \frac{X_{l}}{r_{lj}} > \tau_{l} \right\} \right],
\end{align*}
with
\begin{align*}
\mathbb{E}\left[X^{\text{Rer}}_{i}(\boldsymbol{\tau})\right] &= \sum_{j=1}^{J} p_{j} \left( \mathbb{E}\left[ \min\left\{\frac{X_{i}}{r_{ij}},\tau_{i}\right\}\right] + \mathbb{E}\left[ \frac{X_{l}}{r_{lj}} \mathbbm{1} \left\{ \frac{X_{i}}{r_{ij}} > \tau_{i} \right\} \right] \right).
\end{align*}

Indeed, an arriving job at server pool $i$ has to wait for all the jobs present, and this mean delay is given by the Pollaczek-Khinchin formula for an $M/G/1$ queue. The expected service time requirement of the job before rerouting is equal to $\mathbb{E}\left[ \min\left\{\frac{X_{i}}{r_{ij}},\tau_{i}\right\}\right]$. With probability $p^{\tau}_{i}$ the job is rerouted to server pool $l$. Again the job has to wait for all the jobs present at this pool. Moreover, in this case the service time requirement of the job was $\tau_{i}$ on server pool $i$ and the expected service time requirement is equal to $\mathbb{E}\left[ \frac{X_{l}}{r_{lj}} \mathbbm{1} \left\{ \frac{X_{i}}{r_{ij}} > \tau_{i} \right\} \right]$ on pool $l$.

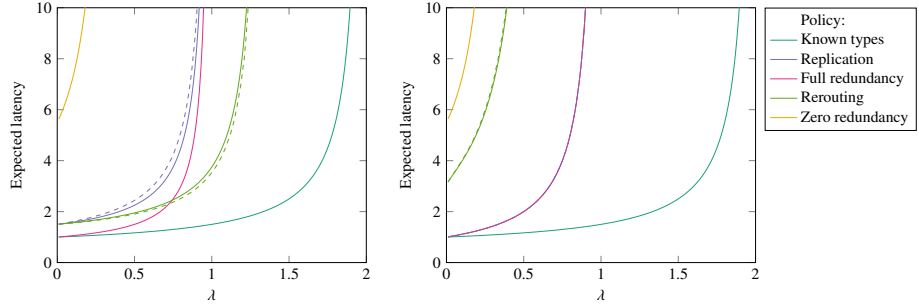
\begin{figure*}[htbp]
  \centering
\resizebox{0.9\textwidth}{!}{\newcommand{\dataFigureA}{TikZFigures/ExpectedLatency_Deg_N2_p05_rslow01.csv}
\newcommand{\dataFigureB}{TikZFigures/ExpectedLatency_Exp_N2_p05_rslow01.csv}

\begin{tabular}{@{}cc@{}}
\begin{tikzpicture}
		\begin{axis}[
			xlabel=$\lambda$,
			ylabel=Expected latency,
			ymin=0,
			ymax=10,
			xmin=0,
			xmax=2,
			no markers]
		\addplot+ table [x=lambda, y=KJ, col sep=comma]{\dataFigureA};
		\pgfplotsset{cycle list shift=1}
		\addplot+ table [x=lambda, y=Rep, col sep=comma]{\dataFigureA};
		\addplot+ table [x=lambda, y=Red, col sep=comma]{\dataFigureA};
		\addplot+ table [x=lambda, y=Rer, col sep=comma]{\dataFigureA};
		\addplot+ table [x=lambda, y=DN, col sep=comma]{\dataFigureA};
		\pgfplotsset{cycle list shift=-3}
		\addplot+[dashed] table [x=lambda, y=RepApprox, col sep=comma]{\dataFigureA};
		\pgfplotsset{cycle list shift=-2}		
		\addplot+[dashed] table [x=lambda, y=RerApprox, col sep=comma]{\dataFigureA};
		\end{axis}
\end{tikzpicture}
&
\begin{tikzpicture}
		\begin{axis}[
			xlabel=$\lambda$,
			ylabel=Expected latency,
			ymin=0,
			ymax=10,
			xmin=0,
			xmax=2,
			no markers,
			legend pos= outer north east,
			legend cell align=left]
		\addlegendimage{empty legend}
		\addplot+ table [x=lambda, y=KJ, col sep=comma]{\dataFigureB};
		\pgfplotsset{cycle list shift=1}
		\addplot+ table [x=lambda, y=Rep, col sep=comma]{\dataFigureB};
		\addplot+ table [x=lambda, y=Red, col sep=comma]{\dataFigureB};
		\addplot+ table [x=lambda, y=Rer, col sep=comma]{\dataFigureB};
		\addplot+ table [x=lambda, y=DN, col sep=comma]{\dataFigureB};
		\pgfplotsset{cycle list shift=-3}
		\addplot+[dashed] table [x=lambda, y=RepApprox, col sep=comma]{\dataFigureB};
		\pgfplotsset{cycle list shift=-2}
		\addplot+[dashed] table [x=lambda, y=RerApprox, col sep=comma]{\dataFigureB};
		
		\addlegendentry{Policy:}
		\addlegendentry{Known types}
		\addlegendentry{Replication}
		\addlegendentry{Full redundancy}
		\addlegendentry{Rerouting}
		\addlegendentry{Zero redundancy}
		\end{axis}
\end{tikzpicture}
\end{tabular}}
\caption{Expected latency for identical replicas in the scenario $I=2$, $J=2$, $N=2$, $n_{1}=n_{2}=1$, $(r_{11},r_{21})=(1,0.1)$ and $(r_{12},r_{22})=(0.1,1)$ with corresponding probabilities $p_{1}=p_{2}=0.5$ when varying the arrival rate $\lambda$ and degenerate (left) and exponential (right) service times. The approximations are depicted by the dashed lines. For degenerate job sizes, the replication policy is depicted with fixed $\boldsymbol{\tau}=\boldsymbol{1}$, while $\boldsymbol{\tau}=\boldsymbol{0}$ gives a lower expected latency. For exponential job sizes, the replication and full redundancy policy are overlapping.}
\label{fig: expected latency N2 rslow01}
\end{figure*}

The expected latency of a job that is initially assigned to server pool $i$ under the $S(A,\boldsymbol{\tau})$ replication policy is given by
\begin{align*}
\mathbb{E}\left[T_{i}^{\text{Rep}}(A,\boldsymbol{\tau})\right] \approx \frac{\lambda \mathbb{E}\left[\left(B^{\text{Rep}}_{i}(A,\boldsymbol{\tau})\right)^{2}\right]}{2\left(1- \lambda \mathbb{E}\left[B^{\text{Rep}}_{i}(A,\boldsymbol{\tau})\right]\right)} + \mathbb{E}\left[X^{\text{Rep}}_{i}(\boldsymbol{\tau})\right],
\end{align*}
where, for $i=1,2$, the expected service time requirement is given by Equation~\eqref{eq: replication service requirement pool i} and
\begin{align*}
\mathbb{E}\left[\left(B^{\text{Rep}}_{i}(A,\boldsymbol{\tau})\right)^{2} \right] &= \sum_{j=1}^{J} p_{j} \alpha_{ij} \bigg( \mathbb{E}\left[ \left( \min\left\{\frac{X_{i}}{r_{ij}},\tau_{i}\right\} \right)^{2} \right] + k^{2}_{il,j}(\tau_{i}) \bigg) + \sum_{j=1}^{J} p_{j} \alpha_{lj} k^{2}_{li,j}(\tau_{l}), 
\end{align*} 
with
\begin{align*}
\mathbb{E}\left[X^{\text{Rep}}_{i}(\boldsymbol{\tau})\right] &= \sum_{j=1}^{J} p_{j} \left( \mathbb{E}\left[ \min\left\{\frac{X_{i}}{r_{ij}},\tau_{i}\right\}\right] + k_{il,j}(\tau_{i}) \right),
\end{align*}  
for $i=1,2$.

Indeed, an arriving job at server pool $i$ has to wait for all the jobs present, and this mean delay is again given by the Pollaczek-Khinchin formula for an $M/G/1$ queue.
The expected service time requirement of the job before replication is equal to $\mathbb{E}\left[ \min\left\{\frac{X_{i}}{r_{ij}},\tau_{i}\right\}\right]$. After replication the (remaining) expected service time requirement is $k_{il,j}(\tau_{i})$.

To obtain the expected latency of an arbitrary job we can simply sum, the expected latency for a job that is initially assigned to server pool $i$ multiplied by the probability of assigning this job to server pool $i$, over $i$. For example, in the rerouting policy we obtain
\begin{align*}
\mathbb{E}\left[T^{\text{Rer}}(A,\boldsymbol{\tau})\right] = \sum_{i=1}^{2} \sum_{j=1}^{J} p_{j} \alpha_{ij} \mathbb{E}\left[T_{i}^{\text{Rer}}(A,\boldsymbol{\tau})\right].
\end{align*}

\subsection*{Numerical results}

We provide numerical results to give further insights in the expression derived for the expected latency. 

In Figure~\ref{fig: expected latency N2 rslow01} the expected latency is depicted as a function of $\lambda$, for degenerate and exponentially distributed job sizes. The approximations of the expected latency (dashed lines) appear to be reasonably good. Depending on the variability of the job size distribution either the rerouting or replication policy achieves the lowest expected latency. 

\section{Conclusion and suggestions for further research}
\label{sec: conclusion}
We have quantified the effective load per server for a system with two server pools when job types are completely unknown or partly known, but where jobs can either be rerouted or replicated to the other server pool after receiving service for some amount of time. From the numerical results we observed that in most of the scenarios rerouting nor replication increases the achievable stability bound. Moreover, we observed that for balanced server speeds the zero redundancy policy achieves the maximum achievable stability bound. We also observed that decreasing the uncertainty in job types increases the achievable stability bound in a convex manner. 

Topics for further research include:

(i) Extension to multiple server pools. This is more involved than having two server pools, since after rerouting or replication there is the extra choice to which server pool(s). 

(ii) Extensions of the analysis of the expected latency of Section~\ref{sec: improving expected latency}. One could, e.g., use ideas from~\cite{BT-MQNDR,W-ADPQ} to approximate the departure process of a single-server queue. Moreover, note that analytic expressions for the expected latency are lacking in the case of redundancy and thus it is not known what the optimal mean latency is for generally distributed job sizes. The question how uncertainty of job types affects the expected latency remains interesting. 

(iii) Optimization of the expressions for the achievable stability bound under the $S(A,\boldsymbol{\tau})$ policy.

(iv) Extension where the service does carry over. 
In this case the expected service time requirement for the rerouting policy becomes 
\begin{align*}
\mathbb{E}\left[B^{\text{Rer}}_{i}(A,\boldsymbol{\tau})\right] &= \sum_{j=1}^{J} p_{j} \alpha_{ij} \mathbb{E}\left[ \min\left\{\frac{X}{r_{ij}},\tau_{i}\right\}\right] + \sum_{j=1}^{J} p_{j} \alpha_{lj} \mathbb{E}\left[ \frac{X-r_{lj}\tau_{l}}{r_{ij}} \mathbbm{1} \left\{ \frac{X}{r_{lj}} > \tau_{l} \right\} \right], 
\end{align*}
and for the replication policy
\begin{align*}
\mathbb{E}\left[B^{\text{Rep}}_{i}(A,\boldsymbol{\tau})\right] &= \sum_{j=1}^{J} p_{j} \alpha_{ij} \bigg( \mathbb{E}\left[ \min\left\{\frac{X}{r_{ij}},\tau_{i}\right\}\right] + k_{il,j}(\tau_{i}) \bigg) + \sum_{j=1}^{J} p_{j} \alpha_{lj} k_{li,j}(\tau_{l}). 
\end{align*}
where
\begin{align*}
&k_{il,j}(y) = \mathbb{E}\left[\min\left\{\frac{X}{r_{ij}}-y , \frac{X-r_{ij}\tau_{i}}{r_{lj}} \right\} \mathbbm{1}\left\{\frac{X}{r_{ij}} > y \right\} \right].
\end{align*}

\section*{Acknowledgments}
\label{sec: acknowlegdements}
The work in this paper is supported by the Netherlands Organisation for Scientific Research (NWO) through Gravitation grant NETWORKS 024.002.003.

\bibliographystyle{plain}
\bibliography{references}

\begin{thebibliography}{10}

\bibitem{APS-ESM}
M.F. Aktas, P.~Peng, and E.~Soljanin.
\newblock Effective straggler mitigation: When clones should attack and when?
\newblock {\em ACM SIGMETRICS Performance Evaluation Review}, 45(2):12--14,
  2017.

\bibitem{AGSS-ESM}
G.~Ananthanarayanan, A.~Ghodsi, S.~Shenker, and I.~Stoica.
\newblock Effective straggler mitigation: {A}ttack of the clones.
\newblock {\em NSDI'13 Proceedings of the 10th USENIX conference on Networked
  Systems Design and Implementation}, 11:185--198, 2013.

\bibitem{AKGSLSH-ROMRC}
G.~Ananthanarayanan, S.~Kandula, A.~Greenberg, I.~Stoica, Y.~Lu, B.~Saha, and
  E.~Harris.
\newblock Reining in the outliers in map-reduce clusters using mantri.
\newblock {\em OSDI'10 Proceedings of the 9th USENIX conference on Operating
  Systems Design and Implementation}, pages 265--278, 2010.

\bibitem{BM-LHSS}
K.~Bimpikis and M.G. Markakis.
\newblock Learning and hierarchies in service systems.
\newblock {\em Management Science}, 65(3):1268--1285, 2018.

\bibitem{BT-MQNDR}
G.R. Bitran and D.~Tirupati.
\newblock Multiproduct queueing networks with deterministic routing:
  {D}ecomposition approach and the notion of inference.
\newblock {\em Management Science}, 34(1):75--100, 1988.

\bibitem{B-SPQT}
A.A. Borovkov.
\newblock {\em Stochastic Processes in Queueing Theory.}
\newblock Springer, 1976.

\bibitem{DB-TAS}
J.~Dean and L.A. Barroso.
\newblock The tail at scale.
\newblock {\em Communications of the ACM}, 56(2):74--80, 2013.

\bibitem{GHBSW-DSSJS}
K.~Gardner, M.~Harchol-Balter, A.~Scheller-Wolf, and B.~Van Houdt.
\newblock A better model for job redundancy: Decoupling server slowdown and job
  size.
\newblock {\em IEEE ACM Transactions on Networking}, 25(6):3353--3367, 2017.

\bibitem{GS-SDHS}
K.~Gardner and C.~Stephens.
\newblock Smart dispatching in heterogeneous systems.
\newblock {\em ACM SIGMETRICS Performance Evaluation Review}, 47(2):12--14,
  2019.

\bibitem{J-ERT}
G.~Joshi.
\newblock {\em Efficient Redundancy Techniques to Reduce Delay in Cloud
  Systems}.
\newblock PhD thesis, Massachusetts Institute of Technology, 2016.

\bibitem{J-BSC}
G.~Joshi.
\newblock Boosting service capacity via adaptive task replication.
\newblock {\em ACM SIGMETRICS Performance Evaluation Review}, 45(2):9--11,
  2017.

\bibitem{KY-SN}
F.P. Kelly and E.~Yudovina.
\newblock {\em Stochastic Networks.}
\newblock Cambridge University Press, 2014.

\bibitem{KRW-JRMS}
Y.~Kim, R.~Righter, and R.~Wolff.
\newblock Job replication on multiserver systems.
\newblock {\em Advances in Applied Probability}, 41(2):546--575, 2009.

\bibitem{KR-RAGC}
G.~Koole and R.~Righter.
\newblock Resource allocation in grid computing.
\newblock {\em Journal of Scheduling}, 11:163--173, 2008.

\bibitem{RBB-ASR}
Y.~Raaijmakers, S.C. Borst, and O.J. Boxma.
\newblock Achievable stability in redundancy scheduling.
\newblock {\em Paper in preparation}, 2020.

\bibitem{S-OROQ}
A.L. Stolyar.
\newblock Optimal routing in output-queued flexible server systems.
\newblock {\em Probability in the Engineering and Informational Sciences},
  19(2):141--189, 2005.

\bibitem{VGMSRS-LLR}
A.~Vulimiri, P.B. Godfrey, R.~Mittal, J.~Sherry, S.~Ratnasamy, and S.~Shenker.
\newblock Low latency via redundancy.
\newblock {\em CoNEXT'13 Proceedings of the 9th ACM conference on Emerging
  Networking Experiments and Technologies}, pages 283--294, 2013.

\bibitem{W-IQN}
J.~Walrand.
\newblock {\em An Introduction to Queueing Networks.}
\newblock Prentice Hall, 1988.

\bibitem{W-ADPQ}
W.~Whitt.
\newblock Approximations for departure processes and queues in series.
\newblock {\em Naval Research Logistics Quarterly}, 31(4):499--521, 1984.

\end{thebibliography}

\newpage

\appendix
\section{Additional numerical results}
\label{app sec: additional numerical results}
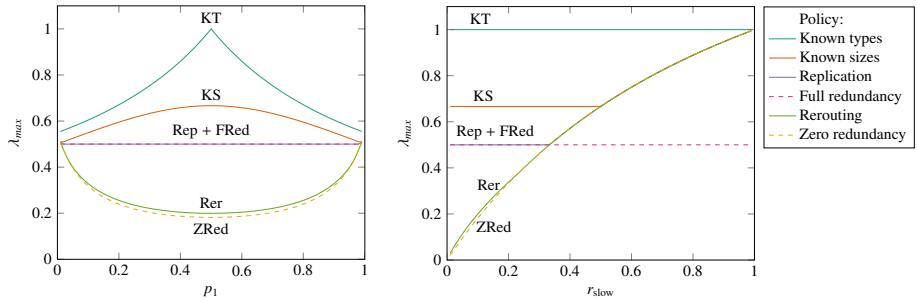
\begin{figure}[htbp]
  \centering
\resizebox{0.9\textwidth}{!}{\newcommand{\dataFigureA}{TikZFigures/StabilityDR_R_TW_SF_Pow_N10_p05_rslow01.csv}
\newcommand{\dataFigureB}{TikZFigures/StabilityDR_R_TW_SF_Pow_N10_p05.csv}

\begin{tabular}{@{}cc@{}}
\begin{tikzpicture}
		\begin{axis}[
			xlabel=$p_{1}$,
			ylabel=$\lambda_{max}$,
			ymin=0,
			ymax=1.1,
			xmin=0,
			xmax=1,
			no markers]
		\addlegendimage{empty legend}
		\addplot+ table [x=prob, y=KT, col sep=comma]{\dataFigureA}
		node[black,pos=0.5,above]{KT};
		\addplot+ table [x=prob, y=KS, col sep=comma]{\dataFigureA}
		node[black,pos=0.5,above]{KS};
		\addplot+ table [x=prob, y=Rep, col sep=comma]{\dataFigureA};
		\addplot+[dashed] table [x=prob, y=Red, col sep=comma]{\dataFigureA}
		node[black,pos=0.5,above]{Rep + FRed};
		\addplot+ table [x=prob, y=Rer, col sep=comma]{\dataFigureA}
		node[black,pos=0.5,above]{Rer};	
		\addplot+[dashed] table [x=prob, y=DN, col sep=comma]{\dataFigureA}
		node[black,pos=0.5,below]{ZRed};		
		\end{axis}
\end{tikzpicture}
&
\begin{tikzpicture}
		\begin{axis}[
			xlabel=$r_{\text{slow}}$,
			ylabel=$\lambda_{max}$,
			ymin=0,
			ymax=1.1,
			xmin=0,
			xmax=1,
			no markers,
			legend pos= outer north east,
			legend cell align=left]
		\addlegendimage{empty legend}
		\addplot+ table [x=rslow, y=KT, col sep=comma]{\dataFigureB}
		node[black,pos=0.1,above]{KT};
		\addplot+ table [x=rslow, y=KS, col sep=comma]{\dataFigureB}
		node[black,pos=0.1,above]{KS};
		\addplot+ table [x=rslow, y=Rep, col sep=comma]{\dataFigureB};
		\addplot+[dashed] table [x=rslow, y=Red, col sep=comma]{\dataFigureB}
		node[black,pos=0.15,above]{Rep + FRed};
		\addplot+ table [x=rslow, y=Rer, col sep=comma]{\dataFigureB}
		node[black,pos=0.25,left]{Rer};	
		\addplot+[dashed] table [x=rslow, y=DN, col sep=comma]{\dataFigureB}
		node[black,pos=0.1,right]{ZRed};		

		\addlegendentry{Policy:}
		\addlegendentry{Known types}
		\addlegendentry{Known sizes}
		\addlegendentry{Replication}
		\addlegendentry{Full redundancy}
		\addlegendentry{Rerouting}
		\addlegendentry{Zero redundancy}
		\end{axis}
\end{tikzpicture}
\end{tabular}}
\caption{Achievable stability bound for identical replicas with Pareto distributed job sizes in the same scenario as Figure~\ref{fig: identical delay stability TwoW_SF Exp N10 p05}, i.e., $I=2$, $J=2$, $N=10$, $n_{1}=n_{2}=5$, $(r_{11},r_{21})=(1,r_{\text{slow}})$ and $(r_{12},r_{22})=(r_{\text{slow}},1)$ with corresponding probabilities $p_{1}$ and $p_{2}=1-p_{1}$, with $r_{\text{slow}}=0.1$ and $p_{1}=p_{2}=0.5$ (right). In the left sub-figure, the replication and full redundancy policy are overlapping.}
\label{app fig: identical delay stability TwoW_SF Pow N10 p05}
\end{figure}

\begin{figure}[htbp]
  \centering
  \resizebox{0.5\textwidth}{!}{\newcommand{\dataFigure}{TikZFigures/StabilityDR_R_TW_SF_Pow_IID_N10_p05.csv}

\begin{tikzpicture}
		\begin{axis}[
			xlabel=$r_{\text{slow}}$,
			ylabel=$\lambda_{max}$,
			ymin=0,
			ymax=6.1,
			xmin=0,
			xmax=1,
			no markers,
			legend pos= outer north east,
			legend cell align=left]
		\addlegendimage{empty legend}
		\addplot+ table [x=rslow, y=KT, col sep=comma]{\dataFigure}
		node[black,pos=0.75,above]{KT};
		\addplot+ table [x=rslow, y=KS, col sep=comma]{\dataFigure}
		node[black,pos=0.75,above]{KS};
		\addplot+ table [x=rslow, y=Rep, col sep=comma]{\dataFigure}
		node[black,pos=0.75,above]{Rep};
		\addplot+[dashed] table [x=rslow, y=Red, col sep=comma]{\dataFigure}
		node[black,pos=0.9,above]{FRed};
		\addplot+ table [x=rslow, y=Rer, col sep=comma]{\dataFigure}
		node[black,pos=0.75,below]{Rer};	
		\addplot+[dashed] table [x=rslow, y=DN, col sep=comma]{\dataFigure}
		node[black,pos=0.5,below]{ZRed};		

		\addlegendentry{Policy:}
		\addlegendentry{Known types}
		\addlegendentry{Known sizes}
		\addlegendentry{Replication}
		\addlegendentry{Full redundancy}
		\addlegendentry{Rerouting}
		\addlegendentry{Zero redundancy}
		\end{axis}
\end{tikzpicture}}
\caption{Achievable stability bound for i.i.d.\ replicas with Pareto distributed job sizes in the same scenario as Figure~\ref{fig: iid delay stability TwoW_SF Exp N10 p05}, i.e., $I=2$, $J=2$, $N=10$, $n_{1}=n_{2}=5$, $(r_{11},r_{21})=(1,r_{\text{slow}})$ and $(r_{12},r_{22})=(r_{\text{slow}},1)$ with corresponding probabilities $p_{1}=p_{2}=0.5$.}
\label{app fig: iid delay stability TwoW_SF Pow N10 p05}
\end{figure}
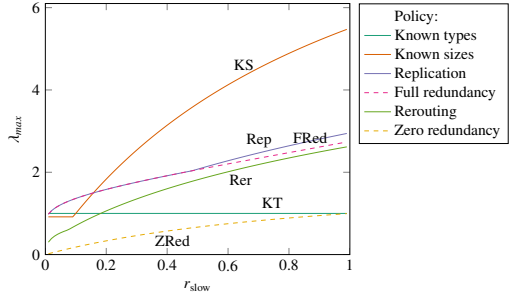

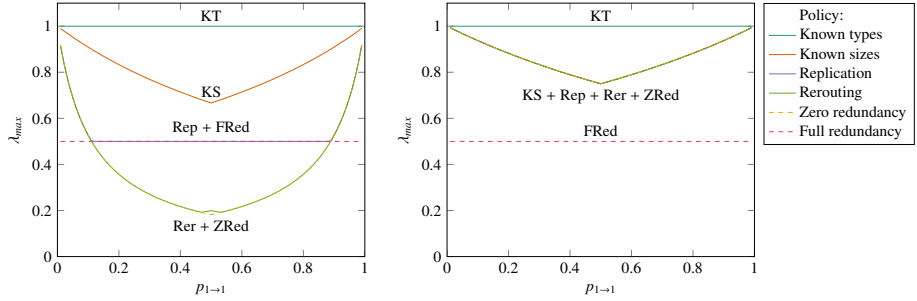
\begin{figure*}[htbp]
\centering
\resizebox{0.9\textwidth}{!}{\newcommand{\dataFigureA}{TikZFigures/StabilityDR_R_TW_SF_Pow_N10_p05_rslow01_likely.csv}
\newcommand{\dataFigureB}{TikZFigures/StabilityDR_R_TW_SF_Pow_N10_p05_rslow06_likely.csv}

\begin{tabular}{@{}cc@{}}
\begin{tikzpicture}
		\begin{axis}[
			xlabel=$p_{1 \rightarrow 1}$,
			ylabel=$\lambda_{max}$,
			ymin=0,
			ymax=1.1,
			xmin=0,
			xmax=1,
			no markers]
		\addplot+ table [x=likely, y=KT, col sep=comma]{\dataFigureA}
		node[black,pos=0.5,above]{KT};
		\addplot+ table [x=likely, y=KS, col sep=comma]{\dataFigureA}
		node[black,pos=0.5,above]{KS};
		\addplot+ table [x=likely, y=Rep, col sep=comma]{\dataFigureA};
		\addplot+[dashed] table [x=likely, y=Red, col sep=comma]{\dataFigureA}
		node[black,pos=0.5,above]{Rep + FRed};
		\addplot+ table [x=likely, y=Rer, col sep=comma]{\dataFigureA};
		\addplot+[dashed] table [x=likely, y=DN, col sep=comma]{\dataFigureA}
		node[black,pos=0.5,below]{Rer + ZRed};		

		\end{axis}
\end{tikzpicture}
&
\begin{tikzpicture}
		\begin{axis}[
			xlabel=$p_{1 \rightarrow 1}$,
			ylabel=$\lambda_{max}$,
			ymin=0,
			ymax=1.1,
			xmin=0,
			xmax=1,
			no markers,
			legend pos= outer north east,
			legend cell align=left]
		\addlegendimage{empty legend}
		\addplot+ table [x=likely, y=KT, col sep=comma]{\dataFigureB}
		node[black,pos=0.5,above]{KT};
		\addplot+ table [x=likely, y=KS, col sep=comma]{\dataFigureB};
		\addplot+ table [x=likely, y=Rep, col sep=comma]{\dataFigureB};
		\pgfplotsset{cycle list shift=1}
		\addplot+ table [x=likely, y=Rer, col sep=comma]{\dataFigureB};
		\addplot+[dashed] table [x=likely, y=DN, col sep=comma]{\dataFigureB}
		node[black,pos=0.5,below]{KS + Rep + Rer + ZRed};
		\addplot+[clr4,dashed] table [x=likely, y=Red, col sep=comma]	{\dataFigureB} node[black,pos=0.5,above]{FRed};		

		\addlegendentry{Policy:}
		\addlegendentry{Known types}
		\addlegendentry{Known sizes}
		\addlegendentry{Replication}
		\addlegendentry{Rerouting}
		\addlegendentry{Zero redundancy}
		\addlegendentry{Full redundancy}
		\end{axis}
\end{tikzpicture}
\end{tabular}}
\caption{Achievable stability bound for identical replicas with Pareto distributed job sizes in the same scenario as Figure~\ref{fig: identical delay stability TwoW_SF Exp N10 rslow01 rslow06 likely}, i.e., $I=2$, $J=2$, $N=10$, $n_{1}=n_{2}=5$, $(r_{11},r_{21})=(1,r_{\text{slow}})$ and $(r_{12},r_{22})=(r_{\text{slow}},1)$ with corresponding probabilities $p_{1}=p_{2}=0.5$, with $r_{\text{slow}}=0.1$ (left) and $r_{\text{slow}}=0.6$ (right) when varying the probability $p_{1 \rightarrow 1}=p_{2 \rightarrow 2}$.}
\label{app fig: identical delay stability TwoW_SF Pow N10 rslow01 likely}
\end{figure*}

\begin{figure*}[htbp]
\centering
\resizebox{0.9\textwidth}{!}{\newcommand{\dataFigureA}{TikZFigures/StabilityDR_R_TW_SF_Pow_IID_N10_p05_rslow01_likely.csv}
\newcommand{\dataFigureB}{TikZFigures/StabilityDR_R_TW_SF_Pow_IID_N10_p05_rslow06_likely.csv}

\begin{tabular}{@{}cc@{}}
\begin{tikzpicture}
		\begin{axis}[
			xlabel=$p_{1 \rightarrow 1}$,
			ylabel=$\lambda_{max}$,
			ymin=0,
			ymax=3.1,
			xmin=0,
			xmax=1,
			no markers]
		\addplot+ table [x=likely, y=KT, col sep=comma]{\dataFigureA}
		node[black,pos=0.5,below]{KT};
		\addplot+ table [x=likely, y=KS, col sep=comma]{\dataFigureA}
		node[black,pos=0.5,above]{KS};
		\addplot+ table [x=likely, y=Rep, col sep=comma]{\dataFigureA};
		\addplot+[dashed] table [x=likely, y=Red, col sep=comma]{\dataFigureA}
		node[black,pos=0.5,above]{Rep + FRed};
		\addplot+ table [x=likely, y=Rer, col sep=comma]{\dataFigureA}
		node[black,pos=0.5,below]{Rer};		
		\addplot+[dashed] table [x=likely, y=DN, col sep=comma]{\dataFigureA}
		node[black,pos=0.5,above]{ZRed};		
		\end{axis}
\end{tikzpicture}
&
\begin{tikzpicture}
		\begin{axis}[
			xlabel=$p_{1 \rightarrow 1}$,
			ylabel=$\lambda_{max}$,
			ymin=0,
			ymax=6.1,
			xmin=0,
			xmax=1,
			no markers,
			legend pos= outer north east,
			legend cell align=left]
		\addlegendimage{empty legend}
		\addplot+ table [x=likely, y=KT, col sep=comma]{\dataFigureB}
		node[black,pos=0.5,above]{KT};
		\addplot+ table [x=likely, y=KS, col sep=comma]{\dataFigureB}
		node[black,pos=0.5,above]{KS};
		\addplot+ table [x=likely, y=Rep, col sep=comma]{\dataFigureB}
		node[black,pos=0.5,above]{Rep};
		\addplot+[dashed] table [x=likely, y=Red, col sep=comma]{\dataFigureB}
		node[black,pos=0.9,below]{FRed};
		\addplot+ table [x=likely, y=Rer, col sep=comma]{\dataFigureB}
		node[black,pos=0.5,below]{Rep};
		\addplot+[dashed] table [x=likely, y=DN, col sep=comma]{\dataFigureB}
		node[black,pos=0.5,below]{ZRed};
		
		\addlegendentry{Policy:}
		\addlegendentry{Known types}
		\addlegendentry{Known sizes}
		\addlegendentry{Replication}
		\addlegendentry{Full redundancy}
		\addlegendentry{Rerouting}
		\addlegendentry{Zero redundancy}
		\end{axis}
\end{tikzpicture}
\end{tabular}}
\caption{Achievable stability bound for i.i.d.\ replicas with Pareto distributed job sizes in the same scenario as Figure~\ref{fig: independent delay stability TwoW_SF N10 rslow01 rslow06 likely}, i.e., $I=2$, $J=2$, $N=10$, $n_{1}=n_{2}=5$, $(r_{11},r_{21})=(1,r_{\text{slow}})$ and $(r_{12},r_{22})=(r_{\text{slow}},1)$ with corresponding probabilities $p_{1}=p_{2}=0.5$, with $r_{\text{slow}}=0.1$ (left) and $r_{\text{slow}}=0.6$ (right) when varying the probability $p_{1 \rightarrow 1}=p_{2 \rightarrow 2}$.}
\label{app fig: independent delay stability TwoW_SF Pow N10 rslow01 rslow06 likely}
\end{figure*}
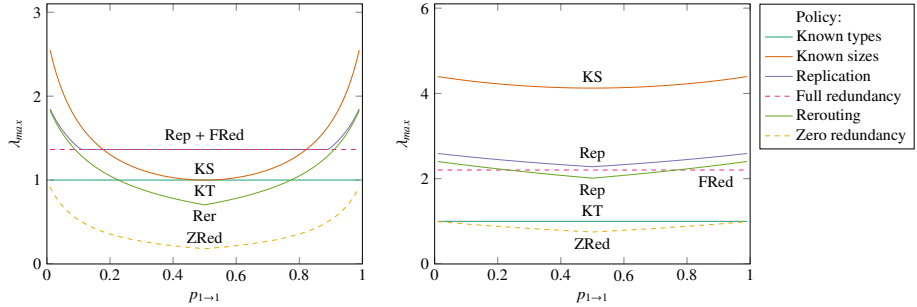

In this appendix we present additional numerical results for the achievable stability bound in case of Pareto Type I distributed job sizes. For Figures~\ref{app fig: identical delay stability TwoW_SF Pow N10 p05}-\ref{app fig: independent delay stability TwoW_SF Pow N10 rslow01 rslow06 likely} the same scenarios as in Figures~\ref{fig: identical delay stability TwoW_SF Exp N10 p05}-\ref{fig: independent delay stability TwoW_SF N10 rslow01 rslow06 likely} are considered, but now for Pareto distributed job sizes, with minimum possible value $1$ and index $\frac{N}{N-1}$ instead of exponentially distributed job sizes with mean $N$.

Figure~\ref{app fig: identical delay stability TwoW_SF Pow N10 p05} reveals that, when comparing the policies for Pareto and exponentially distributed job sizes, the achievable stability bound for the rerouting policy performs worse for Pareto, the other policies achieve the same achievable stability bound.
Only the rerouting policy performs worse for Pareto and (highly) unbalanced server speeds.

Figure~\ref{app fig: iid delay stability TwoW_SF Pow N10 p05} shows that, when comparing the policies for Pareto and exponentially distributed job sizes, the known sizes, replication, full redundancy and rerouting policy all perform better for Pareto distributed job sizes. Similar to Figure~\ref{fig: iid delay stability TwoW_SF Exp N10 p05}, the replication and full redundancy policy both outperform the rerouting policy. 

Figure~\ref{app fig: identical delay stability TwoW_SF Pow N10 rslow01 likely} indicates that, when comparing the policies for Pareto and exponentially distributed job sizes, for unbalanced server speeds, both the replication and rerouting policy perform worse for Pareto distributed job sizes.
For unbalanced server speeds Figure~\ref{app fig: identical delay stability TwoW_SF Pow N10 rslow01 likely} shows that the known sizes, replication, rerouting and zero redundancy policy all achieve the same achievable stability bound. The numerical results for the exponential and Pareto job sizes suggest that these policies are insensitive to the job size distribution, given its mean.

Figure~\ref{app fig: independent delay stability TwoW_SF Pow N10 rslow01 rslow06 likely} reveals that, when comparing the policies for Pareto and exponentially distributed job sizes, the known sizes, replication, full redundancy and rerouting policy all perform better for Pareto distributed job sizes. This is due to the assumption of i.i.d.\ replicas and the heavy tail of the Pareto distribution. 

\section{Load of the system: $n$-threshold policy} 
\label{sec app: learning non threshold policy}

\subsection*{Unknown job sizes}
In this appendix we extend the expressions for the effective load per server, in the case of unknown job sizes, of the rerouting policy, given by Proposition~\ref{thm: stability TwoW_SF_Rer}, by allowing for multiple reroutings (see Definition~\ref{def: n-threshold rerouting policy}). For the rerouting policy this means that after assigning a job to server pool $i$, we can reroute the job to the other pool. In contrast to the threshold policy we can reroute the job back to the pool it was initially assigned to. In the $n$-threshold policy we can in total reroute $n$ times from one pool to the other.\\

As before, let $A$ denote the $2 \times J$ stochastic assignment matrix with elements $\alpha_{ij}$. Let $T$ be the $2 \times n$ rerouting matrix, with rows $(\tau_{i1},\tau_{i2},\dots,\tau_{in})$ denoting the rerouting vector for the job initially assigned to server pool $i$, for $i=1,2$. We define $\tau_{i0}=0$ for $i=1,2$.

\begin{proposition}
The effective load per server in pool~$i$, for $i=1,2$, in the system with rerouting under the $S(A,T)$ policy is
\begin{align*}
\rho_{i,S(A,T)}^{\text{Rer}} = \frac{\lambda\mathbb{E}\left[B^{\text{Rer}}_{i}(A,T)\right]}{n_{i}},
\end{align*}
where,
\begin{align*}
&\mathbb{E}\left[B^{\text{Rer}}_{i}(A,T)\right] = \sum_{j=1}^{J} p_{j} \alpha_{ij} \mathbb{E}\left[ \min\left\{ \frac{X_{i}}{r_{ij}},\tau_{i1} \right\} \right] \\
&\quad + \sum_{m=1}^{n-1}\sum_{j=1}^{J} p_{j} \alpha_{I(m)j} \mathbb{E}\left[ \min\left\{ \frac{X_{i}}{r_{ij}}, \tau_{I(m)m+1} \right\} \right. \\
&\qquad \left. \cdot \mathbbm{1}\left\{\tau_{I(m)m-1} < \frac{X_{i}}{r_{ij}}, \tau_{I(m)m} < \frac{X_{l}}{r_{lj}} \right\} \right] \\
&\quad + \sum_{j=1}^{J} p_{j} \alpha_{I(n)j} \mathbb{E}\left[ \frac{X_{i}}{r_{ij}} \mathbbm{1}\left\{\tau_{I(n)n-1} < \frac{X_{i}}{r_{ij}}, \tau_{I(n)n} < \frac{X_{l}}{r_{lj}} \right\} \right],
\end{align*}        
where $I(m)=l$ if $m$ is odd, and $I(m)=i$ if $m$ is even. 
\end{proposition}

\subsection*{Known job sizes}
\begin{proposition}
The effective load per server in pool~$i$, for $i=1,2$, in the system with rerouting under the $S(A^{x},T^{x})$ policy, in case of known job sizes, is
\begin{align*}
\rho_{i,S(A^{x},T^{x})}^{\text{Rer,KS}} = \frac{\lambda\mathbb{E}\left[B^{\text{Rer,KS}}_{i}(A^{x},T^{x})\right]}{n_{i}},
\end{align*}
where,
\begin{align*}
&\mathbb{E}\left[B^{\text{Rer,KS}}_{i}(A^{x},T^{x})\right] = \iint_{\mathbb{R}^{2}} \left( \sum_{j=1}^{J} p_{j} \alpha^{x}_{ij} \min\left\{ \frac{x_{i}}{r_{ij}}, \tau^{x}_{i1} \right\} \right. \\
&\quad + \sum_{m=1}^{n-1}\sum_{j=1}^{J} p_{j} \alpha^{x}_{I(m)j} \min\left\{ \frac{x_{i}}{r_{ij}}, \tau^{x}_{I(m)m+1} \right\} \\
&\qquad \cdot \mathbbm{1}\left\{\tau^{x}_{I(m)m-1} < \frac{x_{i}}{r_{ij}}, \tau^{x}_{I(m)m} < \frac{x_{l}}{r_{lj}} \right\} \\
&\quad \left. + \sum_{j=1}^{J} p_{j} \alpha^{x}_{I(n)j}  \frac{x_{i}}{r_{ij}} \mathbbm{1}\left\{\tau^{x}_{I(n)n-1} < \frac{x_{i}}{r_{ij}}, \tau^{x}_{I(n)n} < \frac{x_{l}}{r_{lj}} \right\} \right) f_{X}(x_{1},x_{2}) \text{d}x_{1} \text{d}x_{2}.
\end{align*}        
\end{proposition}

\end{document}